\documentclass[12pt]{article}
\usepackage{amssymb}
\usepackage{amsmath}
\newtheorem{thm}{Theorem}
\newtheorem{cor}{Corollary}

\newcommand\R{\mathbb R}
\newcommand\Z{\mathbb Z}
\newcommand\C{\mathbb C}
\newcommand\K{\mathbb K}
\newcommand\N{\mathbb N}
\newcommand\E{\mathbb E}

\def\il{\left\langle}
\def\ir{\right\rangle}
\def\ls{\left\langle}
\def\rs{\right\rangle}
\def\lg{\left|}

\def\eps{\varepsilon}
\def\e{\varepsilon}
\def\g{\gamma}

\def\a{\alpha}
\def\d{\delta}
\def\lall{\Lambda^{\mathop{\rm all}}}
\def\lstd{\Lambda^{\mathop{\rm std}}}

\def\app{{\rm APP}}
\def\intt{{\rm INT}}

\def\lall{\Lambda^{{\rm all}}}
\def\lstd{\Lambda^{{\rm std}}}
\def\e0{e(Q_{0,d})}

\def\ewor{e^{{\rm wor}}}
\def\cwor{{\rm comp}^{{\rm wor}}}
\def\eran{e^{{\rm ran}}}
\def\cran{{\rm comp}^{{\rm ran}}}

\def\cqua{{\rm comp}^{{\rm qua}}}
\def\cqq{{\rm comp}^{{\rm qq}}}
\def\uu{\overline{u}}

\def\C{{\mathbb{C}}}
\newcommand{\Prm}{{\mathbf P}}

\renewcommand{\e}{\varepsilon }
\renewcommand{\epsilon}{\varepsilon }
\renewcommand{\g}{\gamma }
\renewcommand{\rho}{\varrho }

\renewcommand{\phi}{\varphi }
\renewcommand{\a}{\alpha }

\newcommand{\cc}{{\bf c}}

\title{Tractability of Approximation \\
for Weighted Korobov Spaces\\
on Classical and Quantum Computers}
\author{
Erich Novak\thanks{
This work was done while the first and the third authors were visiting
the second author at the University of New South Wales.}
\\
Mathematisches Institut, Universit\"at Jena\\
Ernst-Abbe-Platz 4, 07740 Jena, Germany\\
email: novak@mathematik.uni-jena.de\\
\\
Ian H. Sloan\thanks{The support of the Australian Research Council is greatly
acknowledged.} \\
School of Mathematics, University of New South Wales \\
Sydney 2052, Australia\\
email: i.sloan@unsw.edu.au\\
\\
Henryk Wo\'zniakowski\thanks{
The support of NSF and DARPA is greatly acknowledged.
Effort sponsored by the Defense Advanced Research Projects Agency
(DARPA) and Air Force Research Laboratory, Air Force Materiel Command,
USAF, under agreement number F30602-01-2-0523. The U.S, Government is
authorized to reproduce and distribute reprints for Governmental
purposes notwithstanding any copyright annotation thereon.
The views and conclusions contained herein are those of the authors and
should not be interpreted as necessarily representing the official policies or
endorsements, either expressed or implied, of the Defense Advanced
Research Projects Agency (DARPA), the Air Force Research Laboratory,
or the U.S. Government.}
\\
Department of Computer Science, Columbia University\\
New York, NY 10027, USA, and \\
Institute of Applied Mathematics and Mechanics, University of Warsaw\\
ul. Banacha 2, 02-097 Warszawa, Poland\\
email: henryk@cs.columbia.edu}

\date{May 2002}

\begin{document}
\maketitle

\newpage

\begin{abstract}
We study the approximation problem (or problem of optimal recovery in the
$L_2$-norm) for weighted Korobov spaces with smoothness
parameter $\a$. The weights~$\gamma_j$ of the Korobov spaces moderate
the behavior of periodic functions with respect to successive variables.
The non-negative smoothness parameter $\a$ measures the decay
of Fourier coefficients. For $\a=0$, the Korobov space is the
$L_2$ space, whereas for positive $\a$, the Korobov space
is a space of periodic functions with some smoothness
and the approximation problem
corresponds to a compact operator. The periodic functions are defined on
$[0,1]^d$ and our main interest is when the dimension $d$ varies and
may be large. We consider algorithms using two different classes of information.
The first class $\lall$ consists of arbitrary linear functionals.
The second class $\lstd$ consists of only function values
and this class is more realistic in practical computations.

We want to know when the approximation problem is
tractable. Tractability means that there exists an algorithm whose error
is at most $\e$ and whose information cost is bounded by a polynomial
in the dimension $d$ and in $\e^{-1}$. Strong tractability means that
the bound does not depend on $d$ and is polynomial in $\e^{-1}$.
In this paper we consider the worst case, randomized and quantum
settings. In each setting, the concepts of error and cost are defined
differently, and therefore tractability and strong tractability
depend on the setting and on the class of information.

In the worst case setting, we apply known results to prove
that strong tractability and tractability in the class $\lall$
are equivalent. This holds
iff $\a>0$ and the sum-exponent $s_{\g}$ of weights is finite, where
$s_{\g}\,=\, \inf\big\{\,s>0\ :\ \sum_{j=1}^\infty\g_j^s\,<\,\infty\,\big\}$.

In the worst case setting for the class $\lstd$ we must assume
that $\a>1$ to guarantee that
functionals from $\lstd$ are continuous. The notions of strong
tractability and tractability are not equivalent. In particular,
strong tractability holds iff $\a>1$ and
$\sum_{j=1}^\infty\g_j<\infty$.

In the randomized setting, it is known that randomization does not
help over the worst case setting in the class $\lall$. For the class
$\lstd$, we prove that strong tractability and tractability
are equivalent and this holds under the same assumption
as for the class $\lall$ in the worst case setting, that is,
iff $\a>0$ and $s_{\g} < \infty$.

In the quantum setting, we consider only upper bounds for the class
$\lstd$ with $\a>1$. We prove that $s_{\g}<\infty$ implies strong
tractability.

Hence for $s_{\g}>1$, the randomized and quantum settings
both break worst case intractability of approximation for
the class $\lstd$.

We indicate cost bounds on algorithms with error at
most $\e$. Let $\cc(d)$ denote the cost of computing $L(f)$ for
$L\in \lall$ or $L\in \lstd$, and let the cost of one arithmetic
operation be taken as unity.
The information cost bound in the worst case setting for the
class $\lall$ is of order $\cc (d) \cdot \e^{-p}$
with $p$ being roughly equal to $2\max(s_\g,\a^{-1})$.
Then for the class $\lstd$
in the randomized setting, we obtain the total cost
of order $\cc(d)\,\e^{-p-2} + d\,\e^{-2p-2}$,
 which for small $\e$ is roughly
$$
d\,\e^{-2p-2}.
$$
In the quantum setting, we present a quantum algorithm
with error at most $\e$ that
uses about only  $d + \log \e^{-1}$ qubits
and whose total cost is of order
$$
(\cc(d) +d) \,  \e^{-1-3p/2}.
$$
The speedup of the quantum setting over the randomized setting is of order
$$
\frac{d}{\cc(d)+d}\,\left(\frac1{\e}\right)^{1+p/2}.
$$
Hence, we have a polynomial speedup of order $\e^{-(1+p/2)}$.
We stress that $p$ can be arbitrarily large, and in this case
the speedup is huge.
\end{abstract}

\section{Introduction}

We study the approximation problem (or problem of optimal recovery in the
$L_2$-norm) for periodic functions $f: [0,1]^d \to \C$ that belong to
Korobov spaces.
These are the most studied spaces of periodic functions.
Usually, the unweighted case, in which all variables play the same
role, is analyzed.
As in \cite{HW,SWkor}, in this paper we analyze a more general case
of weighted Korobov spaces,
in which the successive variables
may have diminishing importance. We consider
the unit ball of weighted Korobov spaces $H_d$.
Hence we assume that $\Vert f \Vert_d \le 1$ where the norm depends
on a non-negative smoothness parameter $\a$ and
a sequence $\g=\{\g_j\}$ of positive weights. For $\a=0$ we have
$\|f\|_d=\|f\|_{L_2([0,1]^d)}$, and for $\a>0$ the norm is given by
$$
\|f\|_d\,=\,\bigg(\sum_{h\in \Z^d}r_{\a}(\g,h)\,|\hat f(h)|^2\bigg)^{1/2}
$$
where $\Z^d\,=\,\{\,\dots,-1,0,1,\dots\,\}^d$,
Fourier coefficients are denoted by $\hat f(h)$, and
\begin{equation}\label{rdef}
r_{\a}(\g,h)\,=\,\prod_{j=1}^dr_{\a}(\g_j,h_j)\qquad\mbox{with}\qquad
r_{\a}(\g_j,h_j)\,=\,\left\{\begin{array}{rr} 1\ \ & \mbox{if\ } h_j=0,\\
            \g_j^{-1}|h_j|^{\a} & \mbox{if\ } h_j\not=0,\end{array}\right.
\end{equation}

The smoothness parameter $\a$ measures the decay of the Fourier
coefficients. It is known that the weighted Korobov space $H_d$ consists
of functions that are $k_j$ times differentiable with respect to the $j$th
variable if $k_j\le \a/2$. For $\a\ge0$, the space $H_d$ is a Hilbert
space, and for $\a>1$, it is a Hilbert space with a reproducing
kernel.

The weights $\gamma_j$ of Korobov spaces moderate
the behavior of periodic functions with respect to successive variables.
For $\|f\|_d\le 1$ and for small $\g_j$, we have large $r_{\a}(\g,h)$
with non-zero $h_j$ and therefore the corresponding Fourier
coefficient $|\hat f(h)|$ must be small. In the limiting case when
$\g_j$ approaches zero, all Fourier coefficients $\hat f(h)$ with
non-zero $h_j$ must be zero, that is, the function $f$ does not depend
on the $j$th variable.

We consider algorithms using different classes of information.
We study the two classes $\lall$ and $\lstd$ of information.
The first one $\lall=H_d^*$ consists of all continuous linear
functionals, whereas the second one $\lstd$, called the standard
information, is more realistic in practical computations and consists
only of function values, i.e., of $L_x(f)=f(x)\ \forall f\in H_d$
with $x\in [0,1]^d$. Such functionals are continuous only if $\a>1$.

Our main interest is when the dimension $d$ varies and may be large.
In particular, we want to know when the approximation problem is
tractable. Tractability means that there exists an algorithm
whose error is at most $\e$ and whose information cost (i.e.,
the number of information evaluations from $\lall$ or $\lstd$)
is bounded by a polynomial in the dimension $d$ and in $\e^{-1}$.
Strong tractability means that the bound does not depend on $d$
and is polynomial in $\e^{-1}$. The exponent of strong tractability
is defined roughly as the minimal non-negative
$p$ for which the bound is of order $\e^{-p}$.

We consider the worst case, randomized and quantum settings. Each
setting has its own definition of error, information and total cost.
In the worst
case setting we consider only deterministic algorithms, whose error,
information and total costs are defined by their worst performance.
In the randomized setting
we allow randomized algorithms, and their error and costs are defined on the
average with respect to randomization for a worst function from the
unit ball of $H_d$. In the quantum setting we allow quantum algorithms
that run on a (hypothetical) quantum computer, with the corresponding
definitions of error and costs. Clearly, the concepts of tractability
and strong tractability depend on the setting and on the class
of information. We are interested in checking how the setting
and the class of information change conditions on tractability.

The approximation problem corresponds to the embedding operator
between the weighted Korobov space $H_d$ and the space $L_2([0,1]^d)$.
This operator is compact iff $\a>0$. That is why for $\a=0$ we obtain
negative results in all three settings and for the two classes of
information.

In Section~\ref{det} we study the worst case setting.
It is enough to consider linear algorithms of the form
$$
A_{n,d}(f)\,=\,\sum_{k=1}^na_kL_k(f).
$$
Here, the $a_k$'s are some elements of $L_2([0,1]^d)$,
and the $L_k$'s are some
continuous linear functionals from $\lall$ or $\lstd$.
The functions $a_k$ do not depend on $f$; they form
the fixed output basis of the algorithm.
Necessary and sufficient conditions on tractability of approximation
in the worst case setting easily follow from \cite{HW,WWwei,WWpow}.
With
$$
s_{\g}=\inf\bigg\{s>0:\, \sum_{j=1}^\infty\g_j^s\,<\,\infty\,\bigg\},
$$
we have:
\begin{enumerate}
\item
Let $\a\ge0$.
Strong tractability and tractability of approximation in the class
$\lall$ are equivalent, and this holds iff $\a > 0$ and the
sum-exponent $s_{\g}$ is finite. If so, the exponent of strong tractability is
$$
p^*(\lall)\,=\,2\,\max\left(s_\g,\a^{-1}\right)\
$$

\item Let $\a>1$.
Strong tractability of approximation in the class $\lstd$ holds iff
$$
\sum_{j=1}^\infty \g_j\,<\,\infty.
$$
If so, then $p^*(\lall)\le 2$ and
the exponent of strong tractability $p^*(\lstd)$ satisfies
$$
p^*(\lstd)\,\in \,[p^*(\lall),\,p^*(\lall)+2].
$$

\item Let $\a>1$.
Tractability of approximation in the class $\lstd$ holds iff
$$
a\,:=\,\limsup_{d\to\infty}\frac{\sum_{j=1}^d\g_j}{\ln\,d}\,<\,\infty.
$$
\end{enumerate}
\vskip 1pc
In particular, we see that for the classical unweighted Korobov space,
in which $\g_j=1$ for all $j$, the approximation problem is
intractable. To break intractability we must take
weights $\g_j$ converging to zero
with a polynomial rate, that is,  $\g_j=O(j^{-k})$ for some positive
$k$. Then $s_{\g}\le 1/k$.

In Section~\ref{mc} we study the randomized setting.
We consider randomized algorithms of the form
$$
A_{n,d}(f,\omega)\,=\,\phi_{\omega}\left(L_{1,\omega}(f),L_{2,\omega}(f),
\dots,L_{n,\omega}(f)\right),
$$
where $\omega$ is a random element that is
distributed according to a probability measure $\rho$, and
$L_{k,\omega}\in \Lambda$ with $\phi_{\omega}$ being a mapping
>From $\C^n$ into $L_2([0,1]^d)$.
The randomized error of an algorithm $A_{n,d}$ is defined
by taking the square root of the average value of
$\|f-A_{n,d}(f,\omega)\|^2_{L_2([0,1]^2)}$ with respect to
$\omega$ according to a probability measure $\rho$, and then
by taking the worst case with respect to $f$ from the unit ball of $H_d$.

It is known, see \cite{N92},  that randomization does not help
over the worst case setting for the class $\lall$.
That is why, for the class $\lall$,
tractability and strong tractability in the randomized setting
are equivalent to tractability and strong tractability
in the worst case setting.
For the class $\lstd$ we prove:
\begin{enumerate}
\item
Strong tractability and tractability of approximation
are equivalent, and this holds iff $\a>0$ and $s_\g\,<\,\infty$.
In this case, the exponent of strong tractability is in the interval
$[p^*(\lall),\,p^*(\lall)+2]$, where $p^*(\lall)=2\max(s_\g,\a^{-1})$.

\item
For any $p>p^*(\lall)$, we present an algorithm $A_{n,d}$
with $n$ of order $\eps^{-(p+2)}$
and randomized error at most $\eps$.
Let $\cc(d)$ be the cost of computing one function value, and
let the cost of performing one arithmetic operation be taken as unity.
Then the total cost of the algorithm $A_{n,d}$ is of order
$$
\cc(d)\,\left(\frac1{\eps}\right)^{p+2}\,+\,
d\,\left(\frac1{\eps}\right)^{2p+2}
\quad\forall\,d=1,2,\dots\,,\ \forall\,\e\in (0,1).
$$
Hence, the only dependence on $d$ is through $\cc(d)$ and $d$.
Clearly, if $d$ is fixed and $\e$ goes to zero then the second term
dominates and the total cost of $A_{n,d}$ is of order
$$
d\,\left(\frac1{\eps}\right)^{2p+2}.
$$
\end{enumerate}

The essence of these results is that in the randomized setting
there is no difference between
tractability conditions when we use functionals from $\lall$ or from
$\lstd$. This is especially important when $s_{\g}>1$, since
approximation is then intractable in the worst case setting
for the class $\lstd$
independently of $\a$, and is strongly tractable
in the randomized setting for the class $\lstd$.
Hence for $s_{\g}>1$, randomization breaks
intractability of approximation in the worst case setting for the class
$\lstd$.

In Section~\ref{qc} we study the quantum setting.
We consider quantum algorithms that run on a (hypothetical)  quantum computer.
Our analysis in this section is based on the framework
for quantum algorithms introduced in \cite{He1}
that is relevant for the approximate solution of problems of analysis.

We only consider upper bounds for the class $\lstd$ and
weighted Korobov spaces with $\a >1$ and $s_\g<\infty$.
We present a quantum algorithm with error at most $\e$
whose total cost is of order
$$
\big(  \cc(d) + d\big)  \left(\frac1{\eps}\right)^{1+3p/2}
\quad\forall\,d=1,2,\dots\,,\ \forall\,\e\in (0,1)
$$
with $p\approx p^*(\lall)$ being roughly the exponent of strong
tractability in the worst case setting.

The quantum algorithm uses about $d + \log \e^{-1}$ qubits.
Hence, for moderate $d$ and even for large $\e^{-1}$, the number of qubits
is quite modest. This is especially important, since the number of
qubits will be a limiting resource for the foreseeable future.

It is interesting to compare the results in the quantum setting with the
results in the randomized setting for the class $\lstd$.
The number of quantum queries is of order
$\eps^{-1-3p^*(\lall)/2}$ which is smaller than the corresponding
number $\eps^{-2-p}$ of function values in the randomized setting only if
$p^*(\lall)= 2\max(s_\g,\a^{-1})<2$.
This holds when $s_\g<1$, since $\a>1$ has been
already assumed. However, the number of quantum combinatory
operations is always {\it significantly} smaller than the corresponding number
of combinatory operations in the randomized settings.
If $d$ is fixed and $\e$ goes to zero then
the total cost bound in the randomized setting is of order
$d\e^{-2p-2}$ which is significantly larger than
the total cost bound of order
$(\cc(d)+d)\e^{-1-3p/2}$ in the quantum setting.
This means that
the exponent of $\e^{-1}$ in the cost bound in the quantum setting
is $1+p/2$ less than the exponent in
the randomized setting. We do not know whether our upper bounds for
the quantum computer can be improved.

The speedup of the quantum setting over the randomized setting,
defined as the ratio of the corresponding randomized and quantum
costs, is of order
$$
\frac{d}{\cc(d)+d}\,\left(\frac1{\e}\right)^{1+p/2}.
$$
Hence, we have a polynomial speedup of order $\e^{-(1+p/2)}$.
If $p^*(\lall)$ is close to zero, we may also take $p$ close to zero
and then the speedup is roughly $\e^{-1}$. But $p^*(\lall)$ can be
arbitrarily large. This holds for large $s_\g$.
In this case $p$ is also large and the speedup is huge.

We finish our paper with two appendices. The first is about a general
framework for quantum algorithms and the second contains a proof of the
fact that weighted Korobov spaces are algebras. This fact is crucial for our
upper bounds for quantum algorithms and hence for Theorem~\ref{th7}.

\section{Approximation for Weighted Korobov Spaces}

In this section we define approximation
for periodic functions from the weighted Korobov space $H_d$.
The space $H_d$ is a Hilbert space of complex-valued $L_2$ functions
defined on $[0,1]^d$ that are periodic in each variable with period
$1$. The inner product and norm of $H_d$ are defined as follows.
We take a sequence $\g=\{\g_j\}$ of weights such that
$$
1\,\ge\,\g_1\,\ge\,\g_2\,\ge\,\cdots\,>\,0.
$$
Let $\a\ge 0$.
For $h=[h_1,h_2,\dots,h_d]\in \Z^d$ define
$$
r_{\a}(\g,h)\,=\,\prod_{j=1}^dr_{\a}(\g_j,h_j)\qquad\mbox{with}\qquad
r_{\a}(\g_j,h_j)\,=\,\left\{\begin{array}{rr} 1\ \ & \mbox{if\ } h_j=0,\\
            \g_j^{-s}|h_j|^{\a} & \mbox{if\ } h_j\not=0,\end{array}\right.
$$
where $s=1$ for $\a>0$, and $s=0$ for $\a=0$.
Note that $r_{\a}(\g,h)\ge1$ for all $h\in \Z^d$, and the smallest
$r_{\a}(\g,h)$ is achieved for $h=0$ and has the value $1$.

The inner product in $H_d$ is given by
$$
\il f,g\ir_d\,=\,\sum_{h\in \Z^d}
r_{\a}(\g,h)\,\hat f(h)\,\overline{\hat g(h)},
$$
where $h=(h_1,\dots,h_d)$, and $\hat f(h)$ is the Fourier coefficient
$$
\hat f(h)\,=\,\int_{[0,1]^d}\exp\left(-2\pi i\, h\cdot x\right)\,f(x)\,dx,
$$
with $h\cdot x=h_1x_1+\dots+h_dx_d$.
The inner product in $H_d$ can be also written as
$$
\il f,g\ir_d\,=\,\hat f(0) \overline{\hat g(0)}\,+\,
\sum_{h\in \Z^d, h\not=0}r_{\a}(\g,h)\,\hat f(h)\,\overline{\hat g(h)},
$$
thus the zeroth Fourier coefficient is unweighted. The norm
in $H_d$ is
$$
\|f\|_d\,=\,\bigg(\sum_{h\in \Z^d}r_{\a}(\g,h)\,|\hat f(h)|^2\bigg)^{1/2}.
$$

Note that for $\a=0$ we have $r_{0}(\g,h)\equiv 1$, and
$$
\il f,g\ir_d\,=\,\sum_{h\in\Z^d}\hat f(h)\overline{\hat g(h)}\,=\,
\int_{[0,1]^d}f(x)\overline{g(x)}\,dx.
$$
Hence, in this case
$H_d=L_2([0,1]^d)$ is the space of square integrable
functions. Observe that for any $\a\ge0$ we have $H_d\subset L_2([0,1]^d)$
and
$$
\|f\|_{L_2([0,1]^d)}\,\le\,\|f\|_d\qquad \forall\,f\in H_d.
$$
\vskip 1pc
For $\a>1$, the space $H_d$ is a reproducing kernel Hilbert space,
see \cite{A50,Wahba}. That is,
there exists a function $K_d:[0,1]^d\times [0,1]^d\to \C$, called
the reproducing kernel, such that
$K_d(\cdot,y)\in H_d$ for all $y\in [0,1]^d$, and
$$
f(y)\,=\,\il f,K_d(\cdot,y)\ir_d\qquad \forall\,f\in H_d,\ \forall\,
y\in[0,1]^d.
$$
The essence of the last formula is that the linear functional $L_y(f)=f(y)$
for $f\in H_d$
is continuous and its norm is
$$
\|L_y\|\,=\,K_d^{1/2}(y,y)\qquad \forall\,y\in [0,1]^d.
$$
It is known, see e.g. \cite{SWkor}, that the reproducing kernel $K_d$ is
\begin{equation}\label{kernel}
K_d(x,y)\,=\,\sum_{h\in \Z^d}\frac{\exp\big(2\pi i h\cdot(x-y)\big)}
{r_{\a}(\g,h)}.
\end{equation}
This can be rewritten as
$$
K_d(x,y)\,=\,\,\prod_{j=1}^d
\sum_{h=-\infty}^{\infty}\frac{\exp\left(2\pi i h (x_j-y_j)\right)}
{r_{\a}(\g_j,h)}\,=\,
\prod_{j=1}^d\left(1+2\g_j\sum_{h=1}^\infty\frac{\cos\left(2\pi
h(x_j-y_j)\right)}{h^{\a}}\right).
$$
Hence,  $K_d(x,y)$
depends on $x-y$ and takes only real values. From this we have
$$
K_d(y,y)\,=\, \prod_{j=1}^d\left(1+2\g_j\zeta(\a)\right),
$$
where $\zeta$ is the Riemann zeta function,
$\zeta(\a)=\sum_{h=1}^\infty h^{-\a}$. Hence, $\a>1$ guarantees that
$K_d(y,y)$
is well defined and that $\|L_y\|$ is finite.
\vskip 1pc
We return to the general case for $\a\ge 0$. For $\g_j\equiv1$, the space
$H_d$ is the $L_2$ version of the (unweighted) Korobov
space of periodic functions. For general weights $\g_j$, the
space $H_{d}$ is called a {\em weighted} Korobov space.

We now explain the role of weights $\g_j$.
Take $f\in H_d$ with $\|f\|_d\le 1$. For small values of~$\g_j$
we must have small Fourier coefficients $\hat f (h)$ with $h_j\not=0$.
Indeed, $\|f\|_d\le 1$ implies that $r_{\a}(\g,h)|\hat f(h)|^2\le1$,
and for $h_j\not=0$ this implies that $|\hat f(h)|^2
\le \g_j/|h_j|^{\a}\le\g_j$,
as claimed. Thus, small $\g_j$'s correspond to smoother functions
in the unit ball of $H_d$ in the sense that the Fourier coefficients
$\hat f(h)$ with $h_j\not=0$ must scale like $\g_j^{1/2}$ in order to keep
$\|f\|_{d}\le 1$.

The spaces $H_d$ are related to each other when we vary $d$.
Indeed, it is easy to check that for $d_1\le d_2$ we have
$$
H_{d_1}\,\subseteq\,H_{d_2}\quad\mbox{and}\quad
\|f\|_{d_1}=\|f\|_{d_2}\qquad
\forall\,f\in H_{d_1}.
$$
That is, a function of $d_1$ variables from $H_{d_1}$,
when treated as a function of $d_2$ variables with no dependence
on the last $d_2-d_1$ variables,
also belongs to $H_{d_2}$ with the same norm as in $H_{d_1}$.
This means that we have an increasing sequence of spaces
$H_1\subset H_2\subset\cdots\subset\,H_d$,
and an increasing sequence of the unit balls of $H_d$,
$B_1\subset B_2\subset\cdots\subset B_d$, and $H_{d_1}\cap
B_{d_2}=B_{d_1}$ for $d_1\le d_2$.

So far we assumed that
all weights $\g_j$ are positive. We can also
take zero weights as the limiting case of positive weights when we adopt
the convention that $0/0=0$. Indeed, if one of the weights tends to zero,
say $\g_d\to0$, then $r_{\a}(\g,h)$ goes to infinity for all
$h$ with $h_d\not=0$. Thus to guarantee that $\|f\|_d$ remains finite
we must have $\hat f(h)=0$ for all $h$ with $h_d \ne 0$.
This means that $f$ does not depend
on the $x_d$ coordinate. Similarly, if all the
weights $\g_j$ are zero for $j\ge k$ then a function $f$ from $H_{d}$
does not depend on the coordinates $x_k,x_{k+1},\dots, x_d$.
\vskip 1pc
We are ready to define {\it multivariate approximation} (simply called
approximation) as the operator $\app_d:\,H_d\to L_2([0,1]^d)$ given by
$$
\app_d f\,=\,f.
$$
Hence, $\app_d$ is the embedding from the Korobov space $H_d$ to
the space $L_2([0,1]^d)$.
It is easy to see that $\|\app_d\|=1$; moreover
$\app_d$ is a compact embedding iff $\a>0$.
Indeed, consider the operator $W_d:=\app_d^*\,\app_d:H_d\to H_d$, where
$\app^*_d:L_2([0,1]^d)\to H_d$ is the adjoint operator to $\app_d$.
Then for all $f,g\in H_d$ we have
$$
\il W_df,g\ir_d\,=\,\il \app_d\,f,\app_d\,g\ir_{L_2([0,1]^d)}\,=\,
\il f,g\ir_{L_2([0,1]^d)}.
$$
{} From this we conclude that
$$
W_df_h\,=\, r^{-1}_{\a}(\g,h)\,f_h\qquad \forall\,h\in\Z^d,
$$
where $f_h(x)=\exp\left(2\pi i h\cdot x\right)/r_{\a}^{1/2}(\g,h)$.
We have $\|f_h\|_d=1$ and ${\rm span}(f_h\,:\, h\in \Z^d)$ is dense
in $L_2([0,1]^d)$. This yields that $W_d$ has the form
\begin{equation}\label{wd}
(W_df) (x)\,=\,\sum_{h\in \Z^d}r^{-1}_{\a}(\g,h)\,\hat f(h)
\,\exp\left(2\pi i h\cdot x\right)\qquad
\forall\,f\in H_d,
\end{equation}
where for $\a\in [0,1]$ the convergence of the last series
is understood in the $L_2$ sense.

Thus, $H_d$ has an orthonormal basis consisting of eigenvectors of $W_d$,
and $r^{-1}_{\a}(\g,h)$ is the  eigenvalue of $W_d$ corresponding to $f_h$
for $h\in \Z^d$. Clearly,
$$
\|\app_df\|_{L_2([0,1]^d)}\,=\,\il W_df,f\ir_d^{1/2}\qquad \forall\,f\in H_d,
$$
and therefore, since $W_d$ is self adjoint,
$$
\|\app_d\|\,=\,\|W_d\|^{1/2}\,=\,\left(\max_{h\in \Z^d}
r^{-1}_{\a}(\g,h)\right)^{1/2}\,=\,1.
$$

For $\a=0$ we have $\app_d=W_d$ and both are the identity operator
on $L_2([0,1]^d)$, and therefore they are {\it not} compact. In contrast,
for $\a>0$, the
eigenvalues of $W_d$ go to zero as $|h|=|h_1|+|h_2|+\cdots+
|h_d|$ goes to infinity, and therefore the operator $W_d$ is compact and
$\app_d$ is a compact embedding.

\section{Worst Case Setting}  \label{det}

In this section we deal with tractability of approximation in the worst
case setting. To recall the notion of tractability we proceed as follows.
We approximate $\app_d$ by algorithms\footnote{It is known
that nonlinear algorithms as well as adaptive choice of $L_k$
do not help in decreasing the worst case error, see
e.g., \cite{TWW}.} of the form
$$
A_{n,d}(f)\,=\,\sum_{k=1}^na_kL_k(f).
$$
Here, the $a_k$'s are some elements of $L_2([0,1]^d)$,
and the $L_k$'s are some
continuous linear functionals defined on $H_d$.
Observe that the functions $a_k$ do not depend on $f$, they form
the fixed output basis of the algorithm, see \cite{NW00}.
For all the algorithms in this paper we use the optimal
basis consisting of the eigenvectors of $W_d$.
We assume that $L_k\in \Lambda$, and
consider two classes of
information
$\Lambda$. The first class is $\Lambda=\lall=H_d^*$
which consists of {\em all} continuous linear functionals.
That is, $L\in\lall$ iff there exists $g\in H_d$ such that $L(f)=\il f,g\ir_d$
for all $f\in H_d$. The class $\lall$ is well defined for all $\a\ge 0$.
The second class $\Lambda=\lstd$ is called standard information
and is defined only for
$\a>1$,
$$
\Lambda\,=\,\lstd\,=\,\big\{L_x\,:\ x\in [0,1]^d\ \mbox{with}\
L_x(f)=f(x)\ \forall\,f\in H_d\,\big\}.
$$
Hence, the class $\lstd$ consists of function evaluations. They
are continuous linear functionals since $H_d$ is a reproducing kernel
Hilbert space whenever $\a>1$.

The worst case error of the algorithm $A_{n,d}$ is defined as
$$
\ewor(A_{n,d})\,=\,\sup\big\{\,\|f-A_{n,d}(f)\|_{L_2([0,1]^d)}\,
:\ f\in H_d,\ \|f\|_d\,\le\,1\,\big\}\,=\,
\bigg\|\app_d-\sum_{k=1}^na_kL_k(\cdot)\bigg\|.
$$

Let $\cwor(\eps,H_d,\Lambda)$ be the minimal $n$ for which we can find an
algorithm $A_{n,d}$, i.e., find elements $a_k\in L_2([0,1]^d)$
and functionals $L_k\in \Lambda$,
with worst case error at most $\eps \|\app_d\|$, that is,
$$
\cwor(\eps,H_d,\Lambda)\,=\,\min\big\{\,n\,:\ \exists \ A_{n,d}\
\ \mbox{such that}\ \ \ewor(A_{n,d})\le \eps\,\|\app_d\|\ \big\}.
$$
Observe that in our case $\|\app_d\|=1$ and this represents the initial error
that we can achieve by the zero algorithm $A_{n,d}=0$ without sampling
the function. Therefore $\eps\|\app_d\|=\eps$ can be interpreted as
reducing the initial error by a factor $\eps$. Obviously, it is
only
of interest
to consider $\eps<1$.

This minimal number $\cwor(\eps,H_d,\Lambda)$ of functional evaluations
is closely related to the worst case complexity of
the approximation problem, see e.g., \cite{TWW}.
This explains our choice of notation.

We are ready to define tractability, see \cite{W94}.
We say that approximation is
{\it tractable} in the class $\Lambda$
iff there exist nonnegative numbers $C$, $p$ and $q$ such that
\begin{equation}\label{trac}
\cwor(\eps,H_d,\Lambda)\,\le\,C\,\eps^{-p}\,d^{\,q}\qquad \forall\,
\eps\in(0,1),\ \forall\,d \in \N.
\end{equation}
The essence of tractability is that the minimal number of functional
evaluations is bounded by a polynomial in $\eps^{-1}$ and $d$.

We say that approximation is
{\it strongly tractable} in the class $\Lambda$
iff $q=0$ in (\ref{trac}). Hence, strong tractability means that
the minimal number of functional evaluations has a bound
independent of $d$ and polynomially dependent on $\eps^{-1}$.
The infimum of $p$ in (\ref{trac}) is called the {\it exponent} of
strong tractability and denoted by $p^*=p^*(\Lambda)$.
That is, for any positive $\delta$
there exists a positive $C_{\delta}$ such that
$$
\cwor(\eps,H_d,\Lambda)\,\le\,C_{\delta}\,\eps^{-(p^*+\delta)}
\qquad \forall\, \eps\in(0,1),\ \forall\,d \in \N
$$
and $p^*$ is the smallest number with this property.

Necessary and sufficient conditions on tractability of approximation
in the worst case setting easily follow from \cite{HW,WWwei,WWpow}.
In order to present them we need to recall the notion of the
sum-exponent $s_\g$ of the sequence $\g$, see \cite{WWwei},
which is defined as
\begin{equation}\label{sum-exponent}
s_{\g}\,=\, \inf\bigg\{\,s>0\ :\ \sum_{j=1}^\infty\g_j^s\,<\,\infty\,\bigg\},
\end{equation}
with the convention that the infimum of the empty set is taken as infinity.
Hence, for the unweighted case, $\g_j\equiv 1$, we have $s_\g=\infty$.
For $\g_j=\Theta(j^{-\kappa})$ with $\kappa>0$, we have $s_\g=1/\kappa$.
On the other hand, if $s_{\g}$ is finite then for any positive $\delta$
there exists a positive $M_{\delta}$ such that
$k\,\g_k^{s_{\g}+\delta}\,\le\,\sum_{j=1}^\infty\g_j^{s_{\g}+\delta}\,\le\,
M_{\delta}$.
Hence, $\g_k=O(k^{-1/(s_{\g}+\delta)})$. This shows that $s_{\g}$ is finite
iff $\g_j$ goes to zero polynomially fast in $j^{-1}$, and the reciprocal of
$s_{\g}$ roughly measures the rate of this convergence.

We begin with the class $\lall$. Complexity and optimal algorithms
are well known in this case, see e.g., \cite{TWW}.
Let us define
\begin{equation}\label{seth}
R(\eps,d)\,=\,\big\{\,h\in Z^d\,:\, r^{-1}_{\a}(\g,h)\,>\,\eps^{2}\,\big\}.
\end{equation}
as the set of indices $h$ for which the
eigenvalues of $W_d$, see (\ref{wd}), are greater than $\eps^{2}$.
Then the complexity $\cwor(\eps,H_d,\lall)$ is equal to the cardinality
of the set $R(\eps,d)$,
\begin{equation}\label{7plus}
\cwor(\eps,H_d,\lall)\,=\,\big| R(\eps,d)\big|,
\end{equation}
and the algorithm
\begin{equation}\label{algop}
A_{n,d}(f)(x)\,=\,\sum_{h\in R(\eps,d)}\hat f(h)\,\exp\left(2\pi i h\cdot
x\right)
\end{equation}
with $n=\big| R(\eps,d)\big|$ is optimal and has worst case error at most
$\eps$. This simply means that the truncation of the Fourier series
to terms corresponding to the largest eigenvalues of $W_d$ is the best
approximation of the function $f$.

For $\a=0$  all eigenvalues of $W_d$ have the value $1$. Thus for
$\eps< 1$ we have infinitely many eigenvalues
greater than $\eps^2$ even for~$d=1$.
Therefore the cardinality of the set $R(\eps,1)$ and the complexity
are infinite, which means that approximation is not even solvable,
much less tractable.
For $\a>0$ and $d=1$, we obtain
$$
\cwor(\eps,H_1,\lall)\,\approx\,2\,\g_1^{1/\a}\,\eps^{-2/\a}.
$$

It is proven in \cite{WWwei}
that strong tractability and tractability are equivalent,
and this holds
iff $s_{\g}$ is finite. Furthermore, the exponent of strong tractability
is $p^*(\lall)=2\max\left(s_{\g},\a^{-1}\right)$. We stress that the exponent
of strong tractability is determined by the weight sequence $\g$
if $s_{\g}>\a^{-1}$. On the other hand, if $s_{\g}\le\a^{-1}$ then
$p^*(\lall)=2\a^{-1}$, and this exponent appears in the complexity
even when $d=1$. For such weights, i.e., $s_{\g}\le\a^{-1}$,
multivariate
approximation in any number of variables $d$ requires roughly the same
number of functional evaluations as for $d=1$.

We now turn to the class $\lstd$ and assume that $\a>1$.
Formally, tractability of approximation
in the class $\lstd$ has not been studied; however, it is easy
to analyze this problem based on the existing results. First, observe
that approximation is not easier than {\it multivariate integration}
(or simply integration) defined as
$$
\intt_d(f)\,=\,\int_{[0,1]^d}f(x)\,dx\,=\,\hat f(0)\qquad \forall\,f\in H_d.
$$
Indeed, $\|\intt_d\|=1$, and for any algorithm $A_{n,d}(f)
=\sum_{k=1}^na_kf(x_k)$
for some $a_k\in L_2([0,1]^d)$ and some $x_k\in [0,1]^d$, we have
$$
\|\app_df-A_{n,d}(f)\|^2_{L_2([0,1]^d)}\,=\,\sum_{h\in\Z^d}\big|\hat f(h)-
\widehat{A_{n,d}(f)}(h)\big|^2\,\ge\,\bigg|\hat
f(0)-\sum_{k=1}^nb_kf(x_k)\bigg|^2,
$$
with $b_k=\int_{[0,1]^d}a_k(x)\,dx$. Hence, it is not easier to approximate
$\app_d$ than $\intt_d$, and necessary conditions on tractability of
integration are also necessary conditions on tractability for approximation.
It is known, see \cite{HW}, that integration is strongly tractable
iff $\sum_{j=1}^\infty\g_j<\infty$, and is tractable iff
$a:=\limsup_{d\to\infty}\sum_{j=1}^d\g_j/\ln\,d\,<\,\infty$. Hence,
the same conditions are also necessary for tractability of approximation.
Due to \cite{WWpow}, it turns out that these conditions are also sufficient
for tractability of approximation. More precisely, if $\sum_{j=1}^\infty\g_j
<\infty$, then approximation is strongly tractable and its exponent
$p^*(\lstd)\in[p^*(\lall),p^*(\lall)+2]$, see Corollary 2 (i) of \cite{WWpow}.
Clearly, in this case $p^*(\lall)\le 2$.

Assume that $a\in (0,\infty)$. Then there exists a positive $M$
such that
$$
d\,\g_d/\ln\,d\,\le\,\sum_{j=1}^d\g_j/\ln\,d\,<\,M
$$
for all $d$. Hence,
$\g_j=O(j^{-1}\ln\,j)$, and clearly $s_{\g}=1$. Once more, by
Corollary 2 (i) of \cite{WWpow}, we know that for any positive $\delta$
there exists a positive number $C_{\delta}$ such that
the worst case complexity of approximation is bounded by
$C_{\delta}\,\eps^{-(2+\delta)}d^{4\zeta(\a)\,a+\delta}$. This proves
tractability of approximation.
We summarize this analysis in the following theorem.
\begin{thm}

Consider approximation $\app_d\,:\,H_d\,\to\,L_2([0,1]^d)$ in
the worst case setting.
\begin{enumerate}
\item
Let $\a\ge0$.
Strong tractability and tractability of approximation in the class
$\lall$ are equivalent, and this holds iff $s_\g\,<\,\infty$ and
$\a>0$.
In this case,  the exponent of strong tractability is
$$
p^*(\lall)\,=\,2\,\max\left(s_\g,\a^{-1}\right).
$$

\item Let $\a>1$.
Strong tractability of approximation in the class $\lstd$ holds iff
$$
\sum_{j=1}^\infty \g_j\,<\,\infty.
$$
When this holds,
then $p^*(\lall)\le 2$ and the exponent of strong tractability
$$
p^*(\lstd)\in [p^*(\lall),p^*(\lall)+2].
$$

\item Let $\a>1$.
Tractability of approximation in the class $\lstd$ holds iff
$$
a\,:=\,\limsup_{d\to\infty}\frac{\sum_{j=1}^d\g_j}{\ln\,d}\,<\,\infty.
$$
When this holds,
for any positive $\delta$ there exists a positive $C_{\delta}$ such that
$$
\cwor(\eps,H_d,\lstd)\,\le\,C_{\delta}\,
\eps^{-(2+\delta)}\,d^{\,4\zeta(\a)a
+\delta}\qquad\forall\,d=1,2,\dots\,,\ \forall\,\e\in (0,1),
$$
where $\zeta$ is the Riemann zeta function.
\end{enumerate}
\end{thm}

\section{Randomized Setting}  \label{mc}

In this section we deal with tractability of approximation
in the randomized setting for the two classes $\lall$ and $\lstd$.
The randomized setting is precisely defined in \cite{TWW}.
Here we only mention that we consider randomized algorithms
$$
A_{n,d}(f,\omega)\,=\,\phi_{\omega}\bigg(L_{1,\omega}(f),L_{2,\omega}(f),
\dots,L_{n,\omega}(f)\bigg),
$$
where $\omega$ is a random element that is
distributed according to a probability measure $\rho$, and
$L_{k,\omega}\in \Lambda$ with $\phi_{\omega}$ being a mapping
>From $\C^n$ into $L_2([0,1]^d)$. The essence of randomized
algorithms is that the evaluations, as well the way they are combined,
may depend on a random element. The primary example of a randomized
algorithm is the standard Monte Carlo for approximating
multivariate integration which is of the form
$$
A_{n,d}(f,\omega)\,=\,\frac1n\,\sum_{k=1}^nf(\omega_k),
$$
where $\omega=[\omega_1,\omega_2,\dots,\omega_n]$ with
independent and uniformly distributed $\omega_k$ over $[0,1]^d$
which requires $nd$ random numbers from $[0,1]$.
In this case, $L_{k,\omega}(f)=f(\omega_k)$ are function values
at random sample points, and $\phi_{\omega}(y_1,y_2,\dots,y_n)=
n^{-1}\sum_{k=1}^ny_k$ does not depend on $\omega$ and
is a deterministic mapping.

The randomized error of the algorithm $A_{n,d}$ is defined as
$$
\eran(A_{n,d})\,=\,\sup\left\{\, \E^{1/2}\left(\|f-A_{n,d}(f,\omega)
\|^2_{L_2([0,1]^d)}\right)\ :\ f\in H_d,\, \|f\|_d\,\le\,1\,\right\}.
$$
Hence, we first take the square root of the
average value of the error
$\|f-A_{n,d}(f,\omega)\|^2_{L_2([0,1]^2)}$ with respect to
$\omega$ according to the probability measure $\rho$, and then
take the worst case with respect to $f$ from the unit ball of $H_d$.

Let $\cran(\eps,H_d,\Lambda)$ be the minimal $n$ for which we can find an
algorithm $A_{n,d}$, i.e., a measure $\rho$, functionals $L_{k,\omega}$
and a mapping $\phi_{\omega}$, with randomized error at most
$\eps$. That is,
$$
\cran(\eps,H_d,\Lambda)\,=\,\min\big\{\,n\,:\ \exists \ A_{n,d}\
\ \mbox{such that}\ \ \eran(A_{n,d})\le \eps\ \big\}.
$$
Then tractability in the randomized setting is defined
as in
the paragraph containing
(\ref{trac}), with the replacement of $\cwor(\eps,H_d,\Lambda)$
by $\cran(\eps,H_d,\Lambda)$.

We are ready to discuss tractability in the randomized setting
for the class $\lall$. It is proven in \cite{N92} that randomization
does not really help for approximating linear operators
over Hilbert space for the class $\lall$ since
$$
\cwor(2^{1/2}\eps,H_d,\lall)\,\le\,
\cran(\eps,H_d,\lall)\,\le\,
\cwor(\eps,H_d,\lall),
$$
and these estimates hold for all $\eps\in (0,1)$ and for all $d \in \N$.

This means that tractability in the randomized setting
is equivalent to tractability in the worst case setting, and we can
use the first part of Theorem 1 to characterize
tractability also in the randomized setting.

We now turn to the class $\lstd$. It is well known that randomization
may significantly help for some problems.
The most known  example is the standard Monte Carlo for
multivariate integration of $d$ variables,
which requires at most $\eps^{-2}$ random
function values if the $L_2$ norm of a function is at most one,
independently of how large $d$ is.

We now show that randomization also helps for
approximation over Korobov spaces, and may even break
intractability of approximation in the worst case setting.
As we shall see, this will be achieved by a randomized algorithm
using the standard Monte Carlo for approximating
the Fourier coefficients corresponding to the largest eigenvalues
of the operator $W_d$ defined by (\ref{wd}).
To define such an algorithm we proceed as follows.

We assume that $\a>1$ so that the class $\lstd$ is well defined.
Without loss of generality we also assume that approximation
is tractable in the class $\lall$, which is equivalent
to assuming that  $s_\g<\infty$.

We know from Section 2 that $R(\eps/2^{1/2},d)$ is the set of
indices $h$ for which the eigenvalues of $W_d$ are greater than
$\eps^{2}/2$, see (\ref{seth}).
We also know that the cardinality of the set $R(\eps/2^{1/2},d)$ is
exactly equal to $\cwor(\eps/2^{1/2},H_d,\lall)$
and that for any positive $\delta$ there
exists a positive $C_{\delta}$ such that
$$
\big| R(\eps/2^{1/2},d)\big|\,=\,\cwor(\eps/2^{1/2},H_d,\lall)\,\le\,
C_{\delta}\,\eps^{-(p^*(\lall)+\delta)}\quad
\quad\forall\,d=1,2,\dots\,,\ \forall\,\e\in (0,1),
$$
with $p^*(\lall)=2\max(s_\g,\a^{-1})$.

We want to approximate
$$
f(x)\,=\,\sum_{h\in\Z^d}\hat f(h)\exp\left(2\pi ih\cdot x\right)
$$
for $f\in H_d$. The main idea of our algorithm is to approximate
the Fourier coefficients $\hat f(h)$ for $h\in R(\eps/2^{1/2},d)$ by
the standard Monte Carlo, whereas the Fourier coefficients $\hat f(h)$
for $h\notin R(\eps/2^{1/2},d)$ are approximated
simply by zero. That is, the algorithm $A_{n.d}$
takes the form
\begin{equation}\label{our1}
A_{n,d}(f,\omega)(x)\,=\,\sum_{h\in R(\eps/2^{1/2},d)}\left(\frac1n
\sum_{k=1}^nf(\omega_k)\exp\left(-2\pi i h\cdot\omega_k\right)\right)\,
\exp\left(2\pi i h\cdot x\right),
\end{equation}
where, as for the standard Monte Carlo,
$\omega=(\omega_1,\omega_2,\dots,\omega_n)$ with
independent and uniformly distributed $\omega_k$ over $[0,1]^d$.

The last formula can be rewritten as
\begin{equation}\label{our2}
A_{n,d}(f,\omega)(x)\,=\,
\frac1n\sum_{k=1}^nf(\omega_k)\left(\sum_{h\in R(\eps/2^{1/2},d)}
\exp\bigg(-2\pi i h\cdot(x-\omega_k)\bigg)\,\right).
\end{equation}
{} From (\ref{our2}) it is clear that the randomized algorithm
$A_{n,d}$ uses $n$ random function values.

We are ready to analyze the randomized error of the algorithm $A_{n,d}$.
First of all observe that
$$
\int_{[0,1]^d}\big|f(x)-A_{n,d}(f,\omega)\big|^2dx\,=\,
\sum_{h\in R(\eps/2^{1/2},d)}\bigg|\hat f(h)-\frac1n\sum_{k=1}^nf(\omega_k)
e^{-2\pi i h\cdot\omega_k}\bigg|^2\,+\,\sum_{h\notin
R(\eps/2^{1/2},d)}|
\hat f(h)|^2.
$$
We now compute the average value of the last formula with respect to $\omega$.
Using the well known formula for the Monte Carlo randomized error we obtain
$$
\sum_{h\in R(\eps/2^{1/2},d)}\frac{\intt_d(|f|^2)-|\hat f(h)|^2}{n}
\,+\,\sum_{h\notin R(\eps/2^{1/2},d)}|\hat f(h)|^2.
$$
Since $\intt_d(|f|^2)=\sum_{h\in\Z^d}|\hat f(h)|^2\le\|f\|_d^2$, and
\begin{eqnarray*}
\sum_{h\notin R(\eps/2^{1/2},d)}|\hat f(h)|^2\,&=&\,
\sum_{h\notin R(\eps/2^{1/2},d)}r_{\a}(\g,h)|\hat
f(h)|^2/r_{\a}(\g,h)\\ &\le&\,
\tfrac12{\eps^2}\,\sum_{h\notin R(\eps/2^{1/2},d)}r_{\a}(\g,h)|\hat
f(h)|^2\,\le\,\tfrac12{\eps^2}\,\|f\|^2_d,
\end{eqnarray*}
the error of $A_{n,d}$ satisfies
$$
{\eran}(A_{n,d})^2\,\le\,\frac{|R(\eps/2^{1/2},d)|}{n}\,+\,\frac{\eps^2}2.
$$
Taking
\begin{equation}\label{ourn}
n\,=\,\frac{2\,|R(\eps/2^{1/2},d)|}{\eps^2}\,=\,O\left(
\eps^{-(2+p^*(\lall)+\delta)}\right)
\end{equation}
we conclude that the error of $A_{n,d}$ is at most $\eps$.
This is achieved for $n$ given by (\ref{ourn}), which does {\it not}
depend on $d$, and which depends polynomially on $\eps^{-1}$
with an exponent that exceeds the exponent
of strong tractability in the class
$\lall$, roughly speaking, by at most two.
This means that approximation is strongly tractable
in the class $\lstd$ under exactly the same conditions as in the class
$\lall$.

We now discuss the total cost of the algorithm $A_{n,d}$.
This algorithm requires
$n$ function evaluations $f(\omega_k)$. Since $\omega_k$ is a vector
with $d$ components, it seems reasonable to assume
that the cost of one such function evaluation depends on $d$
and is, say, $\cc(d)$. Obviously, $\cc(d)$ should not be exponential
in $d$ since
for large $d$ we could not even compute one function value. On the other hand,
$\cc(d)$ should be at least linear in $d$ since our functions may depend
on all $d$ variables. Let us also assume that we can perform
combinatory operations such as arithmetic
operations over complex numbers, comparisons of real numbers,
and evaluations of exponential functions. For simplicity
assume that the cost of one combinatory operation is taken as unity.
Hence, for given $h$ and $\omega_k$, we can compute the inner product
$h\cdot \omega_k$ and then $\exp(-2\pi i h\cdot\omega_k)$ in cost of order
$d$.

The implementation of the algorithm $A_{n,d}$ can be done as follows.
We compute and output
$$
y_h\,=\, \frac1n
\sum_{k=1}^nf(\omega_k)\exp\left(-2\pi i h\cdot\omega_k\right)
$$
for all $h\in R(\eps/2^{1/2},d)$. This is done in cost of order
$$
n\,\cc(d)\,+\,n\,d\,|R(\eps/2^{1/2},d)|.
$$
Knowing the coefficients $y_h$ we can compute the algorithm $A_{n,d}$
at any vector $x\in [0,1]^d$ as
$$
A_{n,d}(f,\omega)(x)\,=\,\sum_{h\in R(\eps/2^{1/2},d)}y_h\,
\exp\left(2\pi i h\cdot x\right)
$$
with cost of order $d\,|R(\eps/2^{1/2},d)|$.
Using the estimates on $|R(\eps/2^{1/2},d)|$ and $n$ given by
(\ref{ourn}),
we conclude that the total cost of the algorithm $A_{n,d}$ is of order
$$
\left(\frac1{\eps}\right)^{p+2}\, \cc(d)\,+\,
\left(\frac1{\eps}\right)^{2p+2}\,d
$$
with $p=p^*(\lall)+\delta$.
Hence, the only dependence on $d$ is through $\cc(d)$ and $d$.
We stress the difference in the exponents of the number of function values
and the number of combinatory operations used by the algorithm $A_{n,d}$.
For a fixed $\eps$ and varying $d$, the first term of the cost
will dominate the second term when $\cc(d)$ grows more than linearly
in $d$. In this case the first exponent $p+2$
determines the total cost of the algorithm $A_{n,d}$. On the other hand,
for a fixed $d$ and $\eps$ tending to zero, the opposite is true, and
the second term dominates the first term of the cost, and the second
exponent $2p+2$ determines the cost of $A_{n,d}$.
We summarize this analysis in the following theorem.

\begin{thm}

Consider approximation $\app_d\,:\,H_d\,\to\,L_2([0,1]^d)$ in
the randomized setting.
\begin{enumerate}
\item
Let $\a\ge0$.
Strong tractability and tractability of approximation in the class
$\lall$ are equivalent, and this holds iff $s_\g\,<\,\infty$ and
$\a>0$. When this holds, the exponent of strong tractability is
$$
p^*(\lall)\,=\,2\,\max\left(s_\g,\a^{-1}\right).
$$

\item
Let $\a>1$.
Strong tractability and tractability
of approximation in the class $\lstd$ are equivalent, and this holds
under the same conditions as in the class $\lall$, that is, iff
$s_\g<\infty$. When this holds, the exponent of strong tractability
$p^*(\lstd)\in [p^*(\lall),p^*(\lall)+2]$.

\item The algorithm $A_{n,d}$ defined by (\ref{our1})
with $n$ given by (\ref{ourn}) of order roughly $\eps^{-(p^*(\lall)+2)}$
approximates $\app_d$
with randomized error at most $\eps$. For any positive
$\delta$ there exists a positive number $K_{\delta}$ such that
the total cost of the algorithm $A_{n,d}$ is bounded by
$$
K_{\delta}\left(
\left(\frac1{\eps}\right)^{p+2}\, \cc(d)\,+\,
\left(\frac1{\eps}\right)^{2p+2}\,d\right)
\ \quad\forall\,d=1,2,\dots\,,\ \forall\,\e\in (0,1),
$$
with $p=p^*(\lall)+\delta$.
\end{enumerate}
\end{thm}
\vskip 1pc

We now comment on the assumption $\a>1$ that is present for the class $\lstd$.
As we know from Section 3, this assumption is necessary to guarantee that
function values are continuous linear functionals and it was essential
when we dealt with the worst case setting. In the randomized setting,
the situation is different since we are using random function values,
and the randomized
error depends only on function values in the average sense.
This means that $f(x)$ does not have to be well defined everywhere,
and continuity of the linear functional $L_x(f)=f(x)$ is irrelevant.
Since for any $\a\ge0$, the Korobov space $H_\a$ is a subset of $L_2([0,1]^d)$,
we can treat $f$ as a $L_2$ function. This means that in the randomized
setting we can consider the class $\lstd$ for all $\a\ge 0$.
\vskip 1pc
\noindent {\bf Remark 1}
This is true only if we allow the use of random numbers
>From $[0,1]$. If we only allow the use of random bits (coin tossing as a
source of randomness) then again we need function values to be
continuous linear functionals, which is guaranteed
by the condition $\a > 1$, see \cite{Nov95} for a formal definition
of such ``restricted'' Monte Carlo algorithms.
We add that it is easy to obtain random bits from a quantum computer
while it is not possible to obtain random numbers from $[0,1]$.
\vskip 1pc

Observe that the algorithm $A_{n,d}$ is well
defined for any $\a\ge0$ since the standard Monte Carlo algorithm
is well defined for functions from $L_2([0,1]^d)$.
Furthermore, the randomized error analysis did not use the fact that
$\a>1$, and is valid for all $\a>0$. For $\a=0$ the analysis breaks
down since $n$ given by (\ref{ourn}) would then be infinite.
Even if we treat functions in the $L_2$ sense tractability requires
that $s_{\g}$ be finite. Indeed,
for $s_\g=\infty$ we must approximate
exponentially\footnote{We follow a convention of complexity theory that if
the function grows faster than polynomial then we say it is
exponential.} many Fourier coefficients which, obviously,
contradicts tractability. We summarize this comment
in the following corollary.

\begin{cor}
Consider approximation $\app_d\,:\,H_d\,\to\,L_2([0,1]^d)$ in
the randomized setting with $\a\in [0,1]$ in the class $\lstd$.
\begin{enumerate}
\item Strong tractability and tractability of approximation
are equivalent, and this holds iff $\a>0$ and $s_\g\,<\,\infty$.
When this holds, the exponent of strong tractability is in the
interval $[p,p+2]$, where $p=p^*(\lall)=2\max(s_\g,\a^{-1})$.

\item The algorithm $A_{n,d}$ defined by (\ref{our1})
with $n$ given by (\ref{ourn}) of order roughly $\eps^{-(p^*(\lall)+2)}$
approximates $\app_d$ with randomized error at most $\eps$.
\end{enumerate}
\end{cor}

The essence of these results is that in the randomized setting
there is no difference between
tractability conditions when we use functionals from $\lall$ and
when we use random function values. This is especially important
when $s_{\g}>1$, since approximation is then
intractable in the worst case setting for the class $\lstd$
independently of $\a$. Thus we have the following corollary.

\begin{cor}
Let $s_\g>1$. For the class $\lstd$, randomization breaks intractability of
approximation in the worst case setting.
\end{cor}

\section{Quantum Setting}  \label{qc}

Our analysis in this section is based on the framework introduced in
\cite{He1} of quantum algorithms for the approximate solution of
problems of analysis.  We refer the reader to the surveys
\cite{EHI}, \cite{S3}, and to the monographs \cite{Gr}, \cite{Nie}, and
\cite{P} for general reading on quantum computation.

This approach is an extension of the framework of information-based complexity
theory (see \cite{TWW} and, more formally, \cite{Nov95})
to quantum computation. It also extends the binary black box model of
quantum computation (see \cite{BBC:98})
to situations where mappings on spaces of functions
have to be computed. Some of the main notions of quantum algorithms
can be found in Appendix 1.
For more details and background discussion we refer to \cite{He1}.

\subsection{Quantum Summation of a Single Sequence}   \label{q3}

We need results about the summation of finite sequences on a quantum
computer. The summation problem is defined as follows.
For $N\in\N$ and $1\le p\le\infty$, let $L_p^N$ denote
the space of all functions
$g:\{0,1,\dots,N-1\}\to \R$, equipped with the norm
$$
\|g\|_{L_p^N}=\left(\frac{1}{N}\sum_{j=0}^{N-1}|g(j)|^p \right)^{1/p}
\ \ \mbox{if}\  p<\infty,\ \ \mbox{and} \ \
\|g\|_{L_\infty^N}=\max_{0\le j\le N-1} |g(j)|.
$$
Define $S_N:L_p^N\to \R$ by
$$
S_N(g) = \frac{1}{N}\sum_{j=0}^{N-1}g(j)
$$  and let
$$
F=\mathcal{B}_p^N:=\{g\in L_p^N \,|\,\  \|g\|_{L_p^N}\le 1\}.
$$
Observe that $S_N(\mathcal{B}_p^N)=[-1,1]$ for all $p$ and $N$.
We wish to compute $A(g,\eps)$ which approximates $S_N(g)$ with error $\eps$
and with probability at least $\tfrac34$.
That is, $A(g,\eps)$ is a random variable
which is computed by a quantum algorithm  such that the inequality
$|S_N(g)-A(g,\eps)|\le \eps$ holds with probability at least $\tfrac34$.
The performance of a quantum algorithm can be summarized by
the number of quantum queries, quantum operations and qubits.
These notions are defined in Appendix 1. Here we only mention that
the quantum algorithm obtains information on the function values $g(j)$
by using only quantum queries.
The number of quantum operations is defined as the total number of bit
operations performed by the quantum algorithm. The number of qubits is
defined as $m$ if all quantum operations are performed in the Hilbert
space of dimension $2^m$. It is important to seek algorithms
that require as small a number of qubits as possible.

We denote by  $e_n^q(S_N,F)$ the minimal error
(in the above sense, of probability $\ge \tfrac34$)
that can be achieved by a quantum algorithm
using only $n$ queries. The query complexity is defined for
$\varepsilon > 0$ by
$$
\cqq (\e, S_N, F) =
\min\{\,n\ |\ \, e^q_n(S_N,F) \le \varepsilon\}.
$$
The total (quantum) complexity $\cqua (\e, S_N, F)$ is defined as
the minimal total cost of a quantum algorithm that solves the summation
problem to within $\eps$.
The total cost of a quantum algorithm is defined by counting
the total number of quantum queries plus quantum operations
used by the quantum algorithm.
Let ${\bf c}$ be the cost of one evaluation of $g(j)$. It is
reasonable to assume that the cost of one quantum query is
taken as ${\bf c}+m$ since $g(j)$'s are computed and $m$ qubits
are processed by a quantum query, see Appendix 1 for more details.

The quantum summation is solved by the Grover search
and amplitude estimation algorithm which can be found in
\cite{G2} and \cite{BHM:00}.  This algorithm enjoys almost minimal
error and will be repetitively used for approximation as we shall see
in Sections 5.2 and 5.3.

Let us summarize the known results about the order of
$e_n^q(S_N,\mathcal{B}_p^N)$ for $p=\infty$ and $p=2$.
The case $p=\infty$ is due to \cite{G2}, \cite{BHM:00}
(upper bounds) and \cite{NW} (lower bounds).
The results in the case $p=2$  are due to \cite{He1}.
Further results for arbitrary $1 \le p \le \infty$ can be also found
in \cite{He1} and \cite{HN2}. In what follows,
by ``log'' we mean the logarithm to the base 2.

\begin{thm}  \label{summation}
There are constants $c_j>0$ for $j\in \{ 1, \dots , 9 \}$
such that for all $n,N\in\N$ with $2< n\le c_1 N$ we have
$$
e_n^q(S_N,\mathcal{B}_\infty^N) \asymp n^{-1}
$$
and
$$
c_2 n^{-1}\le
e_n^q(S_N,\mathcal{B}_2^N)\le c_3 n^{-1}\log^{3/2}n \cdot \log\log n .
$$
For $\e\le\e_0 < \tfrac12$, we have
$$
\cqq (\e, S_N, \mathcal{B}_\infty^N ) \asymp \min (N, \e^{-1} )
$$
and
$$
c_4 \min(N,\eps^{-1})\le
\cqq (\e, S_N, \mathcal{B}_2^N ) \le c_5 \min (N, \e^{-1}
\log^{3/2} \e^{-1} \cdot  \log\log \e^{-1}  ) .
$$
For $N\ge\eps^{-1}$,
the algorithm for the upper bound uses about $\log N$ qubits and
the total complexity is bounded by
$$
c_6 \,{\bf c}\,\eps^{-1}\, \le\,
\cqua (\e,  S_N, \mathcal{B}_\infty^N )\,\le\, c_7\,{\bf c}\,
\e^{-1} \cdot  \log N
$$
and
$$
c_8 \,{\bf c}\,\eps^{-1}\,\le\,
\cqua (\e,  S_N, \mathcal{B}_2^N )\,\le\, c_9\,{\bf c}\,
\e^{-1} \log^{3/2} \e^{-1} \cdot \log\log
\e^{-1} \cdot  \log N.
$$
\end{thm}

So far we required that the error is no larger than $\e$ with
probability at least $\tfrac34$. To decrease the probability of failure
>From $\tfrac14$ to, say,  $e^{-\ell/8}$ one can repeat the algorithm
$\ell$ times and take the median as the final result.
See Lemma 3 of \cite{He1} for details.

We also assumed so far that $\|g\|_{L_p^N}\le 1$. If this bound
is changed to, say, $\|g\|_{L_p^N}\le M$ then it is enough to rescale
the problem and replace $g(j)$ by $g(j)/M$.
Then we multiply the computed result by $M$ and obtain the
results as in the last theorem with $\eps$ replaced by $M \eps$.

\subsection{The Idea of the Algorithm for Approximation}  \label{q2}

The starting point of our quantum algorithm for approximation
is a deterministic algorithm on a classical computer
that is similar to the randomized algorithm given by (\ref{our1}), namely
\begin{equation}    \label{start1}
A_{N,d}(f)(x)\,=\,\sum_{h\in R(\eps/3,d)}\left(\frac1N
\sum_{j=1}^N f(x_j)\exp\left(-2\pi i h\cdot x_j\right)\right)\,
\exp\left(2\pi i h\cdot x\right),
\end{equation}
where the $x_1, \dots , x_N$ come from a suitable deterministic rule,
and $R(\cdot,d)$ is defined by (\ref{seth}).
The error analysis of $A_{N,d}$ will be based on three types of
errors. The first error arises from replacing
the infinite Fourier series by a finite series
over the set $R(\eps/3,d)$; this error is $\e/3$.
The second error is made since we replace the
Fourier coefficients which are integrals by a quadrature formulas
We will choose $N$ and the deterministic rule for
computing $x_j$ in such a way that the combination of these two
errors yields
\begin{equation}   \label{start2}
\Vert A_{N,d} (f) - f \Vert_{L_2([0,1]^d)} \le \tfrac23\,\e \quad
\forall\,f \in H_d, \
\Vert f \Vert_d \le 1 .
\end{equation}
This will be possible
(see (\ref{bound}) below)
if $N$ is, in general,
exponentially large in $d$. This may look like a serious drawback,
but the point is that we do {\it not} need to
exactly compute the sums in (\ref{start1}). Instead, the sums
\begin{equation}    \label{start3}
\left(\frac1N \sum_{j=1}^N f(x_j)\exp\left(-2\pi i
h\cdot x_j\right)\right)_{h \in R(\e/3,d)}
\end{equation}
will be approximately computed by a quantum algorithm whose cost depends
only logarithmically on $N$.  We have to guarantee that this third
(quantum) error is bounded by $\e/3$, with probability
at least $\tfrac34$. As we shall see, $\log N$ will be at most linear in $d$
and polynomial in $\log \eps^{-1}$, which will allow us to have
good bounds on the total cost of the quantum algorithm.
\vskip 1pc
\noindent {\bf Remark 2}
Observe that the $|R(\e/3,d)|$
sums given by (\ref{start3}) depend only on $N$ function values
of $f$, whereas $h$ takes as many values
as the cardinality of the set $R(\eps/3,d)$.
Since each function value costs $\cc(d)$,
and since $\cc(d)$ is usually much larger
than the cost of one combinatory operation, it seems like a good idea
to compute all sums in (\ref{start3}) simultaneously.
We do not know how to do this efficiently on a quantum computer
and therefore compute these sums sequentially.
\vskip 1pc

\subsection{Quantum Summation Applied to our Sequences}  \label{q33}

As outlined in the previous subsection,
for the approximation problem we need to compute $S_N(g_h)$
for several sequences $g_1,g_2,\dots,g_R$ each of length $N$
with $R=|R(\eps/3,d)|$.
We assume that $g_h \in L_p^N$ for $p=2$ or $p=\infty$,
and $\Vert g_h \Vert_p  \le M$.
We now want to compute $A(g_h,\eps)$ on a quantum computer such that
(with $\eps/3$ now replaced by $\eps$)
\begin{equation}  \label{req}
\sum_{h=1}^R |S_N(g_h) - A(g_h,\eps) |^2 \le \e^2
\end{equation}
with probability at least $\tfrac34$.
In our case the sequences $g_h$ are the terms of
(\ref{start3}) and we assume that we can compute
$g_h(j)=f(x_j)\exp\left(-2\pi i h\cdot x_j\right)$.
The cost ${\bf c}$ of computing one function value $g_h(j)$ is now equal to
$\cc(d)+2d+2$, since we can compute $g_h(j)$ using
one evaluation of $f$ and $2d+2$ combinatory operations
needed to compute the inner product $y=h\cdot x_j$ and $f(x_j)\exp(-2\pi iy)$.
The cost of one call of the
oracle is roughly
\begin{equation}   \label{sec1}
\log N + \cc(d) +2d+2,
\end{equation}
since we need about $\log N$ qubits and the cost of computing $g_h$
is $\cc(d)+2d+2$.

This summation problem can be solved by the
Grover search or amplitude amplification algorithm
mentioned in Section 5.1.
To guarantee that the bound (\ref{req}) holds
it is enough to compute an approximation for each component
with error $\delta=\e R^{-1/2}$. We will assume that
\begin{equation}\label{assumN}
M\,\delta^{-1}\,=\,\frac{M\,R^{1/2}}{\eps}\,\le\, N.
\end{equation}
We can satisfy (\ref{assumN})
by computing each $S_N(g_h)$ independently for each $h$.

We begin with the case $p=\infty$.
To compute one sum with error $\d$ with probability at least
$1 - \eta$ we need
roughly $\log \eta^{-1}$ repetitions of the algorithm
and this requires about
$(M/\d) \log \eta^{-1}$ queries. We put
$\eta R = \tfrac14$ to obtain an algorithm
that computes each sum in such a way that (\ref{req}) holds.
Hence we need roughly
$\frac{M \sqrt{R}}{\e} \log R$ queries for each $g_h$.
Together we need  roughly
\begin{equation}   \label{sec2}
R \cdot \frac{M \sqrt{R}}{\e}\cdot \log R
\ \ \mbox{queries.}
\end{equation}

The case $p=2$ is similar and we need roughly
\begin{equation}   \label{sec3}
R \cdot \frac{M \sqrt{R}}{\e}\cdot
\log^{3/2}
\frac{M \sqrt{R}}{\e}\cdot
\log \log
\frac{M \sqrt{R}}{\e}\cdot
\log R \ \ \mbox{queries}.
\end{equation}
The total cost is of order
\begin{eqnarray}   \label{sec4}
\left( \log N + \cc(d)+2d+2 \right) R
\frac{M \sqrt{R}}{\e} \log R &&\ \  \mbox{for}\ p=\infty,\\
\left( \log N + \cc(d)+2d+2 \right)
R \frac{M \sqrt{R}}{\e}
\log^{3/2}
\frac{M \sqrt{R}}{\e}\cdot
\log \log
\frac{M \sqrt{R}}{\e}\cdot
\log R &&\ \ \mbox{for}\ p=2.
\end{eqnarray}

\subsection{Results on Tractability}

We only consider upper bounds for the class $\lstd$ and
weighted Korobov spaces for $\a >1$ and $s_\g<\infty$.
We combine the idea
>From Subsection \ref{q2} together with the upper bounds from
Subsection \ref{q33}. We need estimates for the
numbers $N$, $M$, and $R$.

We know from Section 2 that $R(\eps/3,d)$ is the set of
indices $h$ for which the eigenvalues of $W_d$ are greater than
$\eps^{2}/9$, see (\ref{seth}).
We also know from (\ref{7plus})
that the cardinality of the set $R(\eps/3,d)$ is exactly
equal to $\cwor(\eps/3,H_d,\lall)$ and that for any positive
$\eta$ there exists a positive $C_{\eta}$ such that
$$
R=\big|
R(\eps/3,d)\big|\,=\,\cwor(\eps/3,H_d,\lall)\,\le\,
C_{\eta}\,\eps^{-(p^*(\lall)+\eta)}\quad
\forall\,d=1,2,\dots\,,\ \forall\,\e\in (0,1).
$$

For $f \in H_d$ with
$\Vert f \Vert_d \le 1$ we know that
$$
|f(y)|\, =\, |\ls f,K_d(\cdot,y)\rs|\,\le\, K_d(y,y)^{1/2}  \,=\,
\prod_{j=1}^d\bigg(1+2\g_j\zeta(\a)\bigg)^{1/2} ,
$$
where $\zeta$ is the Riemann zeta function, and hence
$$
|f(y)|  \le \exp \bigg( \zeta(\a )  \sum_{j=1}^d \g_j \bigg).
$$
This means that when $\sum_{j=1}^\infty \g_j < \infty$ we can apply
the results from Section~\ref{q33} with $p=\infty$ and
$M$ independent of $d$ and of order one.

If $\sum_{j=1}^\infty \g_j = \infty$, which happens when $s_\g>1$
and could happen if $s_\g=1$, we use the quantum results for $p=2$ and need
estimates not only for $N$ in (\ref{start3}) but also for
$M$ that  bounds the $L_2^N$-norms of the terms in (\ref{start3}).

We know from Lemma 2 (ii) in \cite{SWkor} that there are lattice rules
$Q_{N,d}(f)=N^{-1}\sum_{j=1}^Nf(x_j)$ with prime $N$ and
$x_j=\{j\,z/N\}$ for some
non-zero integer $z\in[-N/2,N/2]^d$ and with
$\{\cdot\}$ denoting
the fractional part,  for which
\begin{equation}    \label{bound}
\big| \,  \intt_d(f) - Q_{N,d}(f) \,  \big|
\le \frac{\prod_{j=1}^d (1+ 2\g_j)^{1/2} }{\sqrt N} \cdot \Vert f \Vert_d .
\end{equation}
As in Section~\ref{q2}, we have to guarantee
an error $\d=\eps R^{-1/2}=O(\e^{1+ (p^*(\lall)+2)/2})$ for all integrands
$x \mapsto f_h(x)=f(x) \exp(-2\pi i h \cdot x)$ with $h \in
R(\e/3,d)$. For these integrands $f_h$ we have
\begin{eqnarray*}
\|f_h\|_d^2\,&=&\,\sum_{j\in \Z^d}|\hat f(h+j)|^2r_{\a}(\g,j)=
\sum_{j\in \Z^d}|\hat f(h+j)|^2r_{\a}(\g,h+j)\,\frac{r_{\a}(\g,j)}
{r_{\a}(\g,h+j)}\\
&\le&\,\bigg(\sum_{j\in \Z^d}|\hat f(h+j)|^2r_{\a}(\g,h+j)\bigg)\,
\,\max_{j\in\Z^d}\frac{r_{\a}(\g,j)}{r_{\a}(\g,h+j)}\\
&=&\,\|f\|_d^2\,\max_{j\in\Z^d}\frac{r_{\a}(\g,j)}{r_{\a}(\g,h+j)}.
\end{eqnarray*}

We now show that
\begin{equation}\label{rrr}
\frac{r_{\a}(\g,j)}{r_{\a}(\g,h+j)}\,\le\,
r_{\a}(\g,h)\,\prod_{m=1}^d\max(1,\g_m2^{\a})\qquad
\forall\,j,h\in \Z^d.
\end{equation}
Indeed, since $r_{\a}$ is a product, it is enough to check
(\ref{rrr}) for all components of $r_{\a}$. For the $m$th component
it is easy to check that
$$
\frac{r_{\a}(\g_m,j_m)}{r_{\a}(\g_m,h_m+j_m)}\,\le\,
\max(1,\g_m2^{\a})r_{\a}(\g_m,h_m),
$$
>From which (\ref{rrr}) follows.

In our case $s_\g<\infty$ which implies that $\g_m$ tends to zero
and therefore $\prod_{m=1}^\infty\max(1,\g_m2^{\a})$ is finite.
Furthermore, for $h\in R(\eps/3,d)$
we have $r_\a(\g,h)\le 9/\eps^2$.
Hence, $\Vert f_h \Vert_d =O(1/\e)$ for all $h\in R(\eps/3,d)$.
We replace $\g_j$ by $1$ in (\ref{bound}) and have
$$
\big| \,  \intt_d(f_h) - Q_{N,d}(f_h) \,  \big| =
O\left(\frac{3^{d/2}}{\eps \sqrt{N}}
\right) = O(\eps^{1+(p^*(\lall)+\eta)/2})
$$
if we take $N$ at least of order
$$
N \,\asymp\, 3^d\,\left( \frac{1}{\e} \right)^{4+ p^*(\lall)+\eta}
$$
or
$$
\log N \,\asymp\, d\, +\ \log \e^{-1}.
$$

To bound $M$ we need to consider the $L_2^N$-norms of the terms
$f_h(x_j)=g_h(j)$ in (\ref{start3}).
Since the Korobov space $H_d$ is an algebra,
see Appendix 2,
we know that $|f_h|^2\in H_d$ and
$$
\Vert \, |f_h|^2 \, \Vert_d \,\le\, C(d) \cdot \Vert f_h \Vert_d^2\,=\,
O\left(C(d)\,\e^{-2}\right),
$$
where $C(d)$ is given in Appendix 2.
Applying the bound (\ref{bound}) to the function
$|f_h|^2$, we obtain a bound, in the $L_2^N$-norm,
of the sequence $z_h = (g_h(j))_{j=1, \dots , N}
= (f_h(x_j))_{j=1, \dots , N}$.
This is the number $M$ that we need in our estimates.
We obtain
$$
\Vert z_h \Vert_{L_2^N}^2\,\le\, M^2\, =\,  \intt_d(|f_h|^2) +
O\left(3^{d/2}\,C(d) \, \e^{-2} \,N^{-1/2}\right).
$$
Obviously,
$$
\intt_d(|f_h|^2)\,=\,\widehat{|f_h|^2}(0)\,=\,
\sum_{j\in \Z^d}|\hat f(h+j)|^2\,\le\,
\|f\|_d^2\,\le 1\quad \forall\,h \in \Z^d.
$$
To guarantee that $M$ does not depend on $d$ and is of order $1$,
we take $N$ such
that
$$
\log N \,\asymp\, d + \log C(d) + \log \e^{-1}\, \asymp\, d+ \log \e^{-1},
$$
since $\log C(d)$ is of order $d$ due to Appendix 2.

Putting these estimates together, we obtain estimates
for the quantum algorithm. We use about
$d + \log \e^{-1}$ qubits.
The total cost of the algorithm is of order
$$
(  \cc(d) + d)  \left(\frac1{\eps}\right)^{1+3(p^*(\lall)+\eta)/2} .
$$
Hence, the only dependence on $d$ is through $\cc(d)$ and $d$.
We summarize this analysis in the following theorem.

\begin{thm}  \label{th7}
Consider approximation $\app_d\,:\,H_d\,\to\,L_2([0,1]^d)$ in
the quantum setting with $\a>1$ in the class $\lstd$. Assume that
$s_\g<\infty$.
Then we have strong tractability.
The quantum algorithm solves the problem to within $\eps$
with probability at least $\tfrac34$ and uses
about $d + \log \e^{-1}$ qubits.
For any positive $\d$ there exists a positive number $K_{\delta}$ such that
the total cost of the algorithm is bounded by
$$
K_{\d}\left(
(\cc(d)\,+\, d) \,
\left(\frac1{\eps}\right)^{1+3(p^*(\lall)+\delta)/2}\right)
\ \quad\forall\,d=1,2,\dots\,,\ \forall\,\e\in (0,1).
$$
\end{thm}

\vskip 1pc
It is interesting to compare the results in the quantum setting with the
results in the worst case and randomized settings for the class $\lstd$.
We ignore the small parameter $\d$ in Theorems 1, 2, 4 and 6.
Then if $s_\g>1$, the quantum setting (as well as the randomized setting)
breaks intractability of approximation in the worst case setting
(again for the class $\lstd$).
The number of quantum queries and quantum combinatory operations is
of order $\eps^{-1-3p^*(\lall)/2}$, which is smaller than the corresponding
number of function values in the randomized setting only if $p^*(\lall)<2$.
However, the number of quantum combinatory operations is
always significantly smaller than the
corresponding number of combinatory operations
in the randomized settings.

\section{Appendix 1: Quantum Algorithms}

We present a framework for quantum algorithms, see
\cite{He1} for more details.
Let $D$, $K$ be nonempty sets, and let $\mathcal{F}(D,K)$
denote the set of all functions from $D$ to $K$.
Let $\K$, the scalar field, be either
the field of real numbers $\R$ or the field of complex numbers $\C$, and let
$G$ be a normed space with scalar field $\K$.
Let $S:F\to G$ be a mapping, where $F \subset \mathcal{F}(D,K)$.
We approximate $S(f)$ for $f\in F$ by means of quantum computations.
Let $H_1$ be the
two-dimensional complex Hilbert space $\C^2$, with its unit vector
basis $\{e_0,e_1\}$, and let
$$
H_m=H_1\otimes\cdots\otimes H_1
$$
be the $m$-fold tensor product of $H_1$, endowed with the tensor
Hilbert space structure. It is convenient to let
$$\Z[0,N) := \{0,\dots,N-1\}$$
for $N\in\N$ (as usual, $\N= \{1,2,\dots \}$ and $\N_0=\N\cup\{0\})$.
Let $\mathcal{C}_m = \{\lg i\rs:\, i\in\Z[0,2^m)\}$ be the canonical basis of
$H_m$, where  $\lg i \rs$ stands for
$e_{j_0}\otimes\dots\otimes e_{j_{m-1}}$, and $i=\sum_{k=0}^{m-1}j_k2^{m-1-k}$
is the binary
expansion of $i$. Denote the set of unitary
operators on $H_m$ by $\mathcal{U}(H_m)$.

A quantum query  on $F$ is given by a tuple
\begin{equation}
\label{J1}
Q=(m,m',m'',Z,\tau,\beta),
\end{equation}
where $m,m',m''\in \N, m'+m''\le m, Z\subseteq \Z[0,2^{m'})$ is a nonempty
subset, and
$$\tau:Z\to D$$
$$\beta:K\to\Z[0,2^{m''})$$
are arbitrary mappings. Denote $m(Q):=m$, the number of qubits of $Q$.

Given such a query $Q$, we define for each $f\in F$ the unitary operator
$Q_f$ by setting for
$\lg i\rs\lg x\rs\lg y\rs\in \mathcal{C}_m
=\mathcal{C}_{m'}\otimes\mathcal{C}_{m''}\otimes\mathcal{C}_{m-m'-m''}$:
\begin{equation}
\label{AB1}
Q_f\lg i\rs\lg x\rs\lg y\rs=
\left\{\begin{array}{ll}
\lg i\rs\lg x\oplus\beta(f(\tau(i)))\rs\lg y\rs &\quad \mbox {if}
\quad i\in Z,\\
\lg i\rs\lg x\rs\lg y\rs & \quad\mbox{otherwise,}
 \end{array}
\right.
\end{equation}
where $\oplus$ means addition modulo $2^{m''}$.
Hence the query uses $m'$ bits to represent the index $i$ which
is used to define the argument $\tau(i)$ at which the function is
evaluated. We assume that the cost of one evaluation of $f$ is
${\bf c}$. The value of $f(\tau(i))$ is then coded by the mapping
$\beta$ using $m''$ bits. Usually, the mapping $\beta$
is chosen in a such a way that the $m''$ most significant bits
of $\beta(f(\tau(i)))$ are stored.
The number of bits that are processed is $m'+m''\le m$, and
usually $m'+m''$ is insignificantly less than $m$. That is why we
define the cost of one query as $m+\cc$.

A quantum algorithm  on $F$  with no measurement is a tuple
$A=(Q,(U_j)_{j=0}^n)$,
where $Q$ is a quantum query on $F$, $n\in\N_0$  and
$U_{j}\in \mathcal{U}(H_m)\,(j=0,\dots,n)$, with $m=m(Q)$.
Given $f\in F$,
we let $A_f\in \mathcal{U}(H_m)$ be defined as
\begin{equation}
\label{B1a}
A_f = U_n Q_f U_{n-1}\dots U_1 Q_f U_0.
\end{equation}
We denote by $n_q(A):=n$ the number of queries and by $m(A)=m=m(Q)$ the
number of qubits of $A$. Let $(A_f(x,y))_{x,y\in\mathcal{C}_{m}}$
be the matrix of the
transformation $A_f$ in the canonical basis $\mathcal{C}_{m}$,
$A_f(x,y)=\il x|A_f|y\ir$.

A quantum algorithm on $F$ with output
in $G$ (or shortly, from $F$ to $G$) with $k$ measurements  is a tuple
$$
A=((A_\ell)_{\ell=0}^{k-1},(b_\ell)_{\ell=0}^{k-1},\varphi),
$$
where $k\in\N,$ and $A_\ell\,(\ell=0,\dots,k-1)$ are quantum algorithms
on $F$ with no measurements,
$$
b_0\in\Z[0,2^{m_0}),
$$
for $1\le \ell \le k-1,\,b_\ell$ is a function
$$
b_\ell:\prod_{i=0}^{\ell-1}\Z[0,2^{m_i}) \to \Z[0,2^{m_\ell}),
$$
where we denoted  $m_\ell:=m(A_\ell)$, and $\varphi$ is a function
$$
\varphi:\prod_{\ell=0}^{k-1}\Z[0,2^{m_\ell}) \to G
$$
with values in $G$.
The output of $A$ at input $f\in F$ will be a probability
measure $A(f)$ on $G$,
defined as follows: First put
\begin{eqnarray}
p_{A,f}(x_0,\dots, x_{k-1})&=&
|A_{0,f}(x_0,b_0)|^2 |A_{1,f}(x_1,b_1(x_0))|^2\dots\nonumber\\
&&\dots |A_{k-1,f}(x_{k-1},b_{k-1}(x_0,\dots,x_{k-2}))|^2.\label{M1}
\end{eqnarray}
Then define $A(f)$ by setting
\begin{equation}
\label{M3}
A(f)(C) =\sum_{\phi(x_0,\dots,x_{k-1})\in C}p_{A,f}(x_0,\dots,
x_{k-1}) \quad \forall \,C\subseteq G.
\end{equation}
We let $n_q(A):=\sum_{\ell=0}^{k-1} n_q(A_\ell)$ denote
the number of queries used by $A$.
For brevity we say $A$ is a quantum algorithm if $A$ is a quantum
algorithm with $k$ measurements for $k\ge 0$.

Informally, such an algorithm $A$
starts with a fixed basis state $b_0$ and function $f$,
and applies in an alternating way unitary
transformations $U_{j}$ (not depending on $f$) and the operator $Q_f$
of a certain query.
After a fixed
number of steps the resulting state is measured, which gives a (random) basis
state, say $\xi_0$. This state is memorized and then transformed
(e.g.,\ by a classical
computation, which is symbolized by $b_1$) into
a new basis state $b_1(\xi_0)$.
This is the starting state to which the
next sequence of quantum operations is applied
(with possibly another query and number of qubits).
The resulting state is again measured,
which gives the (random) basis state $\xi_1$. This state is memorized,
$b_2(\xi_0,\xi_1)$ is computed (classically),
and so on. After $k$ such cycles, we obtain
$\xi_0,\dots,\xi_{k-1}$. Then finally an element
$\varphi(\xi_0,\dots,\xi_{k-1})$ of $G$ is computed (e.g.,
\ again on a classical computer) from the results of all measurements.
The probability measure $A(f)$ is its distribution.

The error
of $A$ is defined as follows: Let $0\le\theta< 1$, $f\in F$, and let
$\zeta$ be any random variable with distribution $A(f)$. Then put
$
e(S,A,f,\theta)=\inf\left\{\varepsilon\,\,|\,\,\Prm\{\|S(f)-\zeta\|>
\varepsilon\}\le\theta
\right\}.
$
Associated with this we introduce
$$
e(S,A,F,\theta)=\sup_{f\in F} e(S,A,f,\theta),
$$
$$
e(S,A,f)=e(S,A,f,\tfrac14),
$$
and
$$
e(S,A,F)=e(S,A,F,\tfrac14) = \sup_{f\in F} e(S,A,f).
$$
Of course one could easily replace here $\tfrac14$ by another positive number
$a < \tfrac12$.
The $n$th minimal query error is defined for $n\in\N_0$ as
$$
e_n^q(S,F)=\inf\{e(S,A,F)\,\,|\,\,A\,\,
\mbox{is any quantum algorithm with}\,\, n_q(A)\le n\}.
$$
This is the minimal error which can be reached using at most $n$ queries.
The quantum query complexity is defined for
$\varepsilon > 0$ by
$$
\cqq (\e, S, F) =
\min\{n_q(A)\,\,|\,\, A\,\,\mbox{is any quantum
algorithm with}\,\, e(S,A,F) \le \varepsilon\}.
$$
The quantities $e_n^q(S,F)$ and $\cqq (\e, S, F)$ are inverse to each
other in the following sense:
For all $n\in \N_0$ and $\varepsilon > 0$,
$e_n^q(S,F)\le \varepsilon$ if and only if
$\cqq (\e_1, S, F) \le n$ for all $\varepsilon_1 > \varepsilon$.
Thus, determining the query complexity is equivalent to determining the
$n$th minimal query error.
The total (quantum) complexity $\cqua (\e, S, F)$ is defined similarly.
Here we count the number of quantum gates that are used by the algorithm;
if function values are needed then we put ${\bf c}$
as the cost of one function evaluation.
>From a practical point of view, the number of available qubits
in the near future will be severely limited.
Hence it is a good idea to present algorithms that only use
a small number of qubits.

\section{Appendix 2: Korobov Spaces are Algebras}

We show that the Korobov space $H_d$ is an algebra for $\alpha>1$.
More precisely, we  prove that if $f,g\in H_{d}$ then $fg\in H_d$ and
\begin{equation}\label{appe1}
\| f\,g\|_{d}\, \le\, C(d)\,\| f\|_{d}\,\|g\|_{d},
\end{equation}
with
$$
C(d)\,=\,2^{\,d\,\max(1,\alpha/2)}\,\prod_{j=1}^d\bigg(1+
2\g_j\zeta(\alpha)\bigg)^{1/2}.
$$
For $f(x)=\sum_{j}\hat f(j)\exp(2\pi i j\cdot x)$ and
$g(x)=\sum_{k}\hat g(k)\exp(2\pi i k\cdot x)$,
with $j$ and $k$ varying through $\Z^d$, we have
$$
f(x)g(x)\,=\,\sum_j\sum_k\hat f(j)\hat g(k)\exp(2\pi i(j+k)\cdot x)\,=\,
\sum_h\left(\sum_j\hat f(j)\hat g(h-j)\right)\exp(2\pi i h\cdot x).
$$
Hence, we need to estimate
$$
\|fg\|_d^2\,=\,\sum_{h}\bigg|\sum_j\hat f(j)\hat g(h-j)\,
r^{1/2}_{\alpha}(\g,h)\bigg|^2.
$$
Observe that
$$
r^{1/2}_{\alpha}(\g_m,h_m)\,\le\,c\,\left(r^{1/2}_{\alpha}(\g_m,k_m)+
r^{1/2}_{\alpha}(\g_m,h_m-k_m)\right)\qquad \forall\,k_m\in \Z,
$$
with $c=2^{\max(0,(\alpha-2)/2)}$. This holds for $h_m=0$ since
$c\ge 1$ and $r_{\a}(\g_m,k_m)\ge 1$,
and is also true for $h_m\not=0$ and $k_m=0$.
For other values of $h_m$ and $k_m$, the inequality is equivalent
to $|h_m|^{\alpha/2}\le c(|k_m|^{\alpha/2}+|h_m-k_m|^{\alpha/2})$ which
holds with $c=1$ for $\alpha/2\le 1$, and
with $c=2^{(\alpha-2)/2}$ for $\alpha/2>1$ by the use of the
standard argument.
Applying this inequality $d$ times we get
$$
r^{1/2}_{\alpha}(\g,h)\,\le\,c^d\,\prod_{m=1}^d\left(
r^{1/2}_{\alpha}(\g_m,k_m)+
r^{1/2}_{\alpha}(\g_m,h_m-k_m)\right)\qquad \forall\,k\in \Z^d.
$$
Let $D=\{1,2,\dots,d\}$ and let $u\subset D$. By $\uu=D-u$ we denote
the complement of $u$. Define
$$
r_{\alpha}(\g,h_u)\,=\,\prod_{m\in u}r_{\alpha}(\g_m,h_m),\qquad
r_{\alpha}(\g,h_{\uu})\,=\,\prod_{m\in \uu}r_{\alpha}(\g_m,h_m).
$$
Then we can rewrite the last inequality as
$$
r^{1/2}_{\alpha}(\g,h)\,\le\,c^d\,\sum_{u\subset D}
r^{1/2}_{\alpha}(\g,k_u)\,r^{1/2}_{\alpha}(\g,h_{\uu}-k_{\uu})
\qquad \forall\,k\in \Z^d.
$$
For $u\subset D$, we define
\begin{eqnarray*}
F_u(x)\,&=&\,\sum_j|\hat f(j)|\,r^{1/2}_{\alpha}(\g,j_u)\,
\exp(2\pi i j\cdot x),\\
G_{\uu}(x)\,&=&\,\sum_j|\hat g(j)|\,r^{1/2}_{\alpha}(\g,j_{\uu})\,
\exp(2\pi i j\cdot x).
\end{eqnarray*}
Observe that $F_u$ and $G_{\uu}$ are well defined
functions in $L_2([0,1]^d)$
since $r_{\alpha}(\g,j_u)
\le r_{\alpha}(\g,j)$ for all $u$ and since
$f$ and $g$ are from $H_d$.
In terms of these functions we see that
\begin{eqnarray*}
\bigg|\sum_j\hat f(j)\hat g(h-j)\,r^{1/2}_{\alpha}(\g,h)\bigg|\,&\le&\,
\sum_j|\hat f(j)|\,|\hat g(h-j)|\,r^{1/2}_{\alpha}(\g,h)\\
&\le&\,c^d\,\sum_{u\subset D}
\sum_j|\hat f(j)|\,r^{1/2}_{\alpha}(\g,j_{u})
|\hat g(h-j)|\,r^{1/2}_{\alpha}(\g,h_{\uu}-j_{\uu})\\
&=&\,c^d\sum_{u\subset D}\sum_j\hat F_u(j)\,\hat G_{\uu}(h-j).
\end{eqnarray*}
Therefore
$$
\|f\,g\|_d^2\,\le\,c^{2d}\,\sum_h\left(\sum_{u\subset D}
\sum_j\hat F_u(j)\,\hat G_{\uu}(h-j)\right)^2.
$$
Since the sum with respect to $u$ has $2^d$ terms, we estimate
the square of the sum of these $2^d$ terms by the sum of the squared terms
multiplied by $2^d$, and obtain
$$
\|f\,g\|_d^2\,\le\,2^dc^{2d}\,\sum_{u\subset D}a_u,
$$
where
$$
a_u\,=\, \sum_h\left(\sum_j\hat F_u(j)\,\hat G_{\uu}(h-j)\right)^2.
$$

We now estimate $a_u$. Each $h$ and $j$ may be written as
$h=(h_u,h_{\uu})$ and $j=(j_u,j_{\uu})$, and therefore
\begin{eqnarray*}
a_u&=&\sum_{h_u}\sum_{h_{\uu}}\left(
\sum_{j_u}\sum_{j_{\uu}}\hat F_u(j_u,j_{\uu})\,\hat G_{\uu}(h_u-j_u,
h_{\uu}-j_{\uu})\right)^2\\
&=&
\sum_{h_u}\sum_{h_{\uu}}\left(
\sum_{j_u}\sum_{j_{\uu}}\hat F_u(h_u-j_u,j_{\uu})\,\hat G_{\uu}(j_u,
h_{\uu}-j_{\uu})\right)^2\\
&=&
\sum_{h_u}\sum_{h_{\uu}}
\sum_{j_u}\sum_{j_{\uu}}
\sum_{k_u}\sum_{k_{\uu}}
\hat F_u(h_u-j_u,j_{\uu})
\hat F_u(h_u-k_u,k_{\uu})
\hat G_{\uu}(j_u,h_{\uu}-j_{\uu})
\hat G_{\uu}(k_u,h_{\uu}-k_{\uu}).
\end{eqnarray*}
Note that
$$
\sum_{h_{\uu}}\hat G_{\uu}(j_u,h_{\uu}-j_{\uu})
\hat G_{\uu}(k_u,h_{\uu}-k_{\uu})\,\le\,G(j_u)G(k_u),
$$
where
$$
G(j_u)\,=\, \left(
\sum_{h_{\uu}}\hat G_{\uu}(j_u,h_{\uu})^2\right)^{1/2}.
$$
Similarly,
$$
\sum_{h_{u}}\hat F_{u}(h_u-j_{u},j_{\uu})\,\hat F_{u}(h_u-k_u,k_{\uu})
\,\le\,F(j_{\uu})F(k_{\uu}),
$$
where
$$
F(j_{\uu})\,=\, \left(
\sum_{h_{u}}\hat F_{u}(h_u,j_{\uu})^2\right)^{1/2}.
$$
We obtain
$$
a_u\,\le\,
\sum_{j_u}\sum_{j_{\uu}}
\sum_{k_u}\sum_{k_{\uu}}F(j_{\uu})F(k_{\uu})G(j_u)G(k_u)
\,=\,\left(\sum_{j_{\uu}}F(j_{\uu})\right)^2
\left(\sum_{k_u}G(k_{u})\right)^2.
$$
Observe that
\begin{eqnarray*}
\left(\sum_{j_{\uu}}F(j_{\uu})\right)^2\,&=&\,
\left(\sum_{j_{\uu}}\left(\sum_{j_u}\hat F_u(j_u,j_{\uu})^2\right)^{1/2}
r^{1/2}_{\alpha}(\g,j_{\uu})r^{-1/2}_{\alpha}(\g,j_{\uu})\right)^2\\
&\le&\,
\sum_{j_{\uu}}\left(\sum_{j_u}\hat F_u(j_u,j_{\uu})^2 r_{\alpha}(\g,j_{\uu})
\right)\left(\sum_{j_{\uu}}r^{-1}_{\alpha}(\g,j_{\uu})\right)\\
&=&\,
\left(\sum_{j_{\uu}}\sum_{j_u}
|\hat f(j_u,j_{\uu})|^2r_{\alpha}(\g,j_{u})r_{\alpha}(\g,j_{\uu})\right)
\left(\sum_{j_{\uu}}r^{-1}_{\alpha}(\g,j_{\uu})\right)\\
&=&\,\left(\sum_j|\hat f(j)|^2r_{\alpha}(\g,j)\right)
\left(\sum_{j_{\uu}}r^{-1}_{\alpha}(\g,j_{\uu})\right)\\
&=&\,\|f\|^2_d\,
\sum_{j_{\uu}}r^{-1}_{\alpha}(\g,j_{\uu}).
\end{eqnarray*}
For the last sum we have
$$
\sum_{j_{\uu}}r^{-1}_{\alpha}(\g,j_{\uu})\,=\,
\prod_{m\in \uu}
\left(1+\g_m\sum_{j\not=0}|j|^{-\alpha}\right)\,=\,
\prod_{m\in \uu}
\left(1+2\g_m\zeta(\alpha)\right).
$$
Similarly,
$$
\left(\sum_{k_u}G(k_u)\right)^2\,\le\,\|g\|_d^2\,\sum_{k_u}
r^{-1}_{\alpha}(\g,k_u)\,=\,\|g\|^2_d\,\prod_{m\in u}\left(1+2\g_m
\zeta(\alpha)\right).
$$
Putting all these estimates together we conclude that
\begin{eqnarray*}
\|f\,g\|^2_d\,&\le&\,2^dc^{2d}\sum_{u\subset D}\|f\|^2_d\,\|g\|^2_d\,
\prod_{m\in \uu}
\left(1+2\g_m\zeta(\alpha)\right)\prod_{m\in u}
\left(1+2\g_m\zeta(\alpha)\right)\\
&=&\,
2^dc^{2d}\sum_{u\subset D}\|f\|^2_d\,\|g\|^2_d\prod_{m=1}^d
\bigg(1+2\g_m\zeta(\alpha)\bigg)\\
&=&\,4^dc^{2d}\prod_{m=1}^d\bigg(1+2\g_m\zeta(\alpha)
\bigg)\,\|f\|^2_d\,\|g\|^2_d,
\end{eqnarray*}
>From which (\ref{appe1}) easily follows.
\vskip 1pc
For the quantum setting, we need to consider the function
$w(x)=f(x)\overline{f}(x)=|f(x)|^2$ for $f\in H_d$. Note that
$\overline{f}$ also belongs to $H_d$ and $\|\overline{f}\|_d=\|f\|_d$,
since $\hat{\overline{f}}(h)=\overline{\hat f(-h)}$
and $r_{\alpha}(\g,h)=r_{\alpha}(\g,-h)$ for all $h\in \Z^d$.
Then (\ref{appe1}) guarantees that $w\in H_d$ and
\begin{equation}\label{appe2}
\big\|\,|f|^2\,\big\|_d\,\le\,C(d)\,\|f\|^2_d
\qquad \forall\,f\in H_d.
\end{equation}

\vskip 2pc
{\bf Acknowledgments. }
We are grateful to Stefan Heinrich,
Anargyros Papageorgiou, Joseph F. Traub,
Greg Wasilkowski, and Arthur Werschulz for valuable remarks.


\vskip 2pc

\begin{thebibliography}{99}

\bibitem{A50}
N. Aronszajn (1950):
Theory of reproducing kernels.
Trans. Amer. Math. Soc. {\bf 68}, 337--404.

\bibitem{BBC:98}
R.~Beals,  H.~Buhrman, R.~Cleve, and  M.~Mosca (1998):
Quantum lower bounds by polynomials.
Proceedings of 39th IEEE FOCS, 352--361, see also
http://arXiv.org/abs/quant-ph/9802049.

\bibitem{BHM:00}
G.~Brassard, P.~H{\o}yer, M.~Mosca, and A.~Tapp (2000):
Quantum amplitude amplification and estimation.
Technical report, http://arXiv.org/abs/quant-ph/0005055.

\bibitem{EHI}
A. Ekert, P. Hayden, and H. Inamori (2000):
Basic concepts in quantum computation.
See http://arXiv.org/abs/quant-ph/0011013.

\bibitem{G1}
L. Grover (1996):
A fast quantum mechanical algorithm for database search.
Proc. 28 Annual ACM Symp. on the Theory of
Computing, 212--219, ACM Press New York.
See also http://arXiv.org/abs/quant-ph/9605043.

\bibitem{G2}
L. Grover (1998):
A framework for fast quantum mechanical algorithms.
Proc. 30 Annual ACM Symp. on the Theory of
Computing, 53--62, ACM Press New York.
See also http://arXiv.org/abs/quant-ph/9711043.

\bibitem{Gr}
J. Gruska (1999):
Quantum Computing.
McGraw-Hill, London.

\bibitem{He1}
Heinrich, S. (2002):
Quantum summation with an application to integration.
J. Complexity {\bf 18}.
See also http://arXiv.org/abs/quant-ph/0105116.

\bibitem{He2}
S.\ Heinrich (2001):
Quantum integration in Sobolev classes.
Preprint.
See also http://arXiv.org/abs/quant-ph/0112153.

\bibitem{HN1}
S. Heinrich and E. Novak (2002): Optimal summation
and integration by deterministic,
randomized, and quantum algorithms.
In: Monte Carlo and Quasi-Monte Carlo Methods 2000.
K.-T. Fang, F. J. Hickernell, H. Niederreiter
(eds.), pp. 50--62, Springer.
See also http://arXiv.org/abs/quant-ph/0105114.

\bibitem{HN2}
S. Heinrich and E. Novak (2001):
On a problem in quantum summation.
Submitted to J. Complexity.
See also http://arXiv.org/abs/quant-ph/0109038.

\bibitem{HW}
F. J. Hickernell and H. Wo\'zniakowski (2001):
Tractability of multivariate integration for periodic functions.
J. Complexity {\bf 17}, 660--682.

\bibitem{NW}
A. Nayak and F. Wu (1999):
The quantum query complexity of approximating
the median and related statistics.
STOC, May 1999, 384--393.
See also http://arXiv.org/abs/quant-ph/9804066.

\bibitem{Nie}
M. A. Nielsen and I. L. Chuang (2000):
Quantum Computation and Quantum Information, Cambridge
University Press.

\bibitem{N92}
E. Novak (1992):
Optimal linear randomized methods for linear operators in
Hilbert spaces.
J. Complexity {\bf 8}, 22--36.

\bibitem{Nov95}
E. Novak (1995):
The real number model in numerical analysis.
J. Complexity {\bf 11}, 57--73.

\bibitem{Nov01}
E. Novak (2001):
Quantum complexity of integration.
J. Complexity {\bf 17}, 2--16.
See also http://arXiv.org/abs/quant-ph/0008124.

\bibitem{NW00}
E. Novak and H. Wo\'zniakowski (2000):
Complexity of linear problems with a fixed output basis.
J. Complexity {\bf 16}, 333--362.

\bibitem{NW01}
E. Novak and H. Wo\'zniakowski (2001):
Intractability results for integration and discrepancy,
J. Complexity {\bf 17}, 388--441.

\bibitem{P}
A. O. Pittenger (1999):
Introduction to Quantum Computing Algorithms.
Birk\-h\"auser, Boston.

\bibitem{S3}
P. W. Shor  (2000):
Introduction to Quantum Algorithms.
See http://arXiv.org/abs/quant-ph/quant-ph/0005003.

\bibitem{SW2}
I. H. Sloan and H. Wo\'zniakowski (1998):
When are quasi-Monte Carlo algorithms efficient for high
dimensional integrals?
J. Complexity {\bf 14}, 1--33.

\bibitem{SWkor}
I. H. Sloan and H. Wo\'zniakowski (2001):
Tractability of multivariate integration for weighted Korobov classes.
J. Complexity {\bf 17}, 697--721.

\bibitem{TWW}
J. F. Traub, G. W. Wasilkowski and H. Wo\'zniakowski (1988):
Information-Based Complexity,
Academic Press, New York.

\bibitem{TW}
J. F. Traub and H. Wo\'zniakowski  (2001):
Path integration on a quantum computer,
submitted for publication.
See also http://arXiv.org/abs/quant-ph/0109113.
\bibitem{Wahba}
G. Wahba (1990):
Spline Models for Observational Data,
SIAM-NSF Regional Conference Series in Appl. Math.,
SIAM, {\bf 59}, Philadelphia.

\bibitem{WWwei}
G. W. Wasilkowski and H. Wo\'zniakowski (1999):
Weighted tensor product algorithms for linear multivariate problems.
J. Complexity {\bf 15}, 402--447.

\bibitem{WWpow}
G. W. Wasilkowski and H. Wo\'zniakowski (2001):
On the power of standard information for weighted approximation.
Found. Comput. Math. {\bf 1}, 417-434, 2001.

\bibitem{W94}
H. Wo\'zniakowski (1994):
Tractability and strong tractability of linear
multivariate problems.
J. Complexity {\bf 10}, 96--128.

\bibitem{W}
H. Wo\'zniakowski (1999):
Efficiency of quasi-Monte Carlo algorithms for high dimensional
integrals.
In Monte Carlo and Quasi-Monte Carlo Methods 1998,
eds. H. Niederreiter and J. Spanier, Springer Verlag, Berlin,
114--136.

\end{thebibliography}
\end{document}